\documentclass[11pt]{article}

\usepackage[a4paper, margin=1in]{geometry}
\usepackage{setspace}
\usepackage{microtype} 

\usepackage[linesnumbered, ruled]{algorithm2e}
\usepackage{booktabs, siunitx}
\usepackage{array}
\usepackage{hyperref}
\usepackage{authblk}

\usepackage{amsmath}
\usepackage{mathrsfs}
\usepackage{amsthm}
\usepackage{amssymb}
\usepackage{bm}

\usepackage{color}

\usepackage{multirow}
\usepackage{float}
\usepackage{graphicx}
\usepackage{placeins}
\usepackage{pgfplots}

\usepackage{csquotes}
\usepackage{booktabs}
\usepackage[justification=centering]{caption}

\usepackage{fullwidth}
\usepackage{tcolorbox}
\usepackage{tikz}

\usepackage[style = authoryear, maxcitenames =2, maxbibnames = 20, uniquename=false]{biblatex}
\title{Delta-Vega Hedging with Deep Reinforcement Learning: From Simulation to Reality\thanks{We acknowledge Scuola Normale Superiore for the provision of market data. We thank the participants at the XIII Bachelier World Congress (University of Bologna, 2026), the XXVII Workshop on Quantitative Finance (University of Bergamo, 2026), and the International Fintech Research Conference (University of Pavia, 2026) for insightful comments and discussions. 
}}

\author[1]{Michele Colombi\thanks{Corresponding author. E-mail: \texttt{michele.colombi@sns.it}}}
\author[2,3]{Marco Di Francesco\thanks{E-mail: \texttt{marco.difrancesco3@unibo.it}, \texttt{marco.difrancesco@leitha.eu}}}
\author[2,4]{Pietro Rossi\thanks{E-mail: \texttt{pietro.rossi3@unibo.it}, \texttt{pietro.rossi@prometeia.it}}}
\affil[1]{Scuola Normale Superiore, Pisa, Italy}
\affil[2]{Department of Statistical Sciences, University of Bologna, Italy}
\affil[3]{Leith\`a (Unipol Group), Bologna, Italy}
\affil[4]{Prometeia S.p.A., Bologna, Italy}
\date{November 24, 2026}

\begin{document}

\maketitle

\begin{abstract}

Hedging an option portfolio with the underlying asset alone leaves the position exposed to changes in volatility, which can be managed by trading another derivative, at higher transaction costs. 
We propose a deep reinforcement learning approach to hedge both price and volatility risk under discrete-time trading and transaction costs, by trading the underlying asset and a second, longer-dated option. We train the agent with Proximal Policy Optimization (PPO) on synthetic data from a stochastic volatility model, and restrict its state to quantities observable in the market, so that the policy learned in simulation applies directly to market data. In simulation, we validate the policy against semi-analytical Delta and Delta-Vega hedging benchmarks. 
PPO matches Delta-Vega's objective at baseline transaction costs and increasingly outperforms it as costs rise, with both two-instrument strategies consistently outperforming the single-instrument Delta hedge.
On historical S\&P 500 index option data, we benchmark our approach against practitioner Black-Scholes Delta and Delta-Vega hedging strategies. The two-instrument strategies roughly halve the profit-and-loss dispersion of the Delta hedge, and PPO attains a lower profit-and-loss volatility than Delta-Vega at comparable transaction costs. PPO holds more of the hedging option than Delta-Vega, in line with the stochastic-volatility model hedge, and its advantage is concentrated in high-volatility episodes.

\end{abstract}

\bigskip

\newpage

\newpage

\section{Introduction}

Dynamic option hedging is a longstanding problem in financial theory and practice. In the canonical Black-Scholes (BS) framework \parencite{BlackScholes, merton_theory_1973}, an optimal hedging strategy can be achieved by continuously rebalancing a portfolio consisting of the underlying asset and a risk-free account.
However, the strategy relies on several assumptions such as the absence of transaction costs, continuous-time trading, and constant volatility. In fact, these  conditions do not hold in financial markets. 
Transaction costs and discrete trading frequency introduce a trade-off between cost and hedging accuracy.
Furthermore, as volatility varies over time, a portfolio hedged with the underlying asset alone remains exposed to volatility risk, which can be managed by trading another hedging instrument.

Deep reinforcement learning (RL) has recently emerged as a powerful tool for dynamic optimization problems in finance, as it relies on fewer model assumptions than classical stochastic control and remains applicable where realistic market models become intractable \parencite{hambly_recent_2023}. In option hedging, RL has been shown to effectively learn hedging strategies that account for market frictions, outperforming classical benchmarks such as the BS Delta hedge \parencite{Buhler, Kolm, Du, caoDeepHedgingDerivatives2021}, Delta being the first partial derivative of the option price with respect to the price of the underlying asset.

Despite this flexibility, RL-based approaches have mostly targeted single-instrument Delta-hedging strategies \parencite{pickard_deep_2023}.
A Delta hedge, however, leaves the portfolio exposed to changes in volatility \parencite{bakshiDeltaHedgedGainsNegative2003}. Market participants measure this exposure through the practitioner Black-Scholes Vega, defined as the first partial derivative of the option price with respect to its implied volatility.
By correcting the BS Delta with a Vega-based adjustment, part of the volatility risk can be hedged with the underlying asset alone \parencite{hull_optimal_2017}.
Under stochastic volatility, however, a complete hedge requires a second instrument, i.e.,\ a tradable derivative, as the underlying asset does not complete the market \parencite{romanoContingentClaimsMarket1997}.
In practice, option traders manage the Vega of their positions by trading liquid plain-vanilla options. Since even the most liquid options carry substantially larger bid-ask spreads than the underlying asset, a cost-aware hedging strategy is particularly relevant in this setting.
Yet, only a few RL-based studies use a second hedging instrument under transaction costs, either benchmarking against Gamma hedging \parencite{armstrongDeepGammaHedging2024a,francoisDeepHedgingOptions2025}, Gamma being the second partial derivative of the option price with respect to the underlying asset price, or, when addressing volatility risk, relying on simulated settings as a proof of concept \parencite{Buhler,cao_gamma_2023}.

We build on this line of work and propose a deep RL approach to jointly hedge price and volatility risk using the underlying asset and a hedging option, both subject to transaction costs, which we validate in a controlled simulated environment and transfer directly to real market data.
We leverage Proximal Policy Optimization (PPO) \parencite{ppo} and train the agent on data generated from the \textcite{Heston} stochastic volatility model.
Following \textcite{Kolm}, we assume a mean-variance objective for the agent, capturing the trade-off between costs and replication. Importantly, we restrict the information set available to the agent to market-observable variables only, namely prices, holdings, and contract specifics. We do not give the agent access to model-specific, latent variables driving the data-generating process, so that the policy learned in simulation can be applied directly to real market data without any further estimation.

Our approach is therefore model-agnostic: the Heston model only generates the training data, and the agent learns a hedging policy from simulated price paths, rather than computing sensitivities from a model. We refer to the learned strategy as \textit{Delta-Vega-like}, since it trades the same two instruments as the model's Delta-Vega hedge, but it is not bound to it once transaction costs are introduced or the policy is applied to market data.

We illustrate our approach in three stages.
First, we validate the agent in a frictionless setting where the Heston model implies an optimal, semi-analytical Delta-Vega hedge.
Second, we introduce transaction costs, under which the model hedge is no longer optimal, and the agent is free to depart from it in pursuit of its objective. We consider different transaction cost scenarios reflecting the different spread profiles of the underlying and the hedging option, and study the agent's performance across a range of risk-aversion levels.  
Finally, we transfer the policy learned in simulation to daily S\&P 500 index option data from 2019 to 2022.
This latter step addresses the application of our RL hedging strategy to real market data, an aspect that has received limited attention compared to simulation-based studies \parencite{mikkila23}. 
A strategy that performs well \textit{in-model}, i.e.\ under the assumptions used to derive it, need not be robust \textit{out-of-model}, against market dynamics that the model fails to capture. The empirical evaluation of a policy optimized in simulation, a practice known in the broader machine learning literature as \textit{sim-to-real transfer} \parencite{zhaoSimtorealTransferDeep2020}, is therefore a key step in assessing the applicability of RL hedging strategies in practice.

Our work makes two main contributions.
First, we extend RL hedging to a joint Delta-Vega-like policy that trades the underlying asset and a second, longer-dated option, both subject to transaction costs. The agent is trained on Heston data and observes market variables only. In the absence of costs, it reproduces the semi-analytical Heston Delta-Vega hedge. Under moderate and high costs, it departs from the model hedge and attains a higher mean-variance objective. 
We perform a sensitivity analysis by training the agent on a grid of cost and risk-aversion levels, and find that it adaptively allocates its hedging between the two instruments as costs and risk aversion vary. Both the agent and the Delta-Vega benchmark outperform the Delta-only hedge, even though the option leg is the more expensive one to trade.
Second, we illustrate the applicability of our Delta-Vega-like RL policy by training the agent on Heston data generated from several calibrated parameter sets and maturities, and transferring the learned policy to S\&P 500 index option data from 2019 to 2022. 
We rely on the practitioner Black-Scholes Delta and Delta-Vega hedging as benchmarks, which are widely adopted across market participants,
and find that the PPO agent achieves lower P\&L volatility at comparable transaction costs. 
We observe a systematic over-exposure in the hedging option of the RL policy compared to the practitioner Delta-Vega strategy, in line with the Heston model hedge on which the agent is trained, and find that the agent's advantage over the practitioner benchmark is concentrated in high-volatility episodes.
This highlights the importance of the training environment in shaping the learned policy, and provides an informative account of the applicability of similar deep RL approaches in real settings.

The remainder of the paper is organized as follows. In Section \ref{sec:related_works}, we present an overview of related works. In Section \ref{sec:RL} we introduce the reinforcement learning notation and the PPO algorithm.
In Section \ref{sec:setup} we outline the option hedging setup and show how we embed it into an RL framework.
Section \ref{sec:results} presents numerical results, starting from the synthetic environment case, both with and without transaction costs, and then studying the empirical transfer of the PPO policy to S\&P 500 market data. Finally, Section \ref{sec:conclusion} concludes.

\section{Related Works}\label{sec:related_works}

\textcite{Halperin} is among the first to frame option hedging as a reinforcement learning problem, establishing a formal connection between the BS pricing framework and Q-learning, one of the foundational RL algorithms.
More relevant to this work, \textcite{Kolm} show that RL-based hedging policies, trained to optimize a mean-variance objective under transaction costs and no knowledge of the market model, outperform the BS-Delta strategy.
\textcite{Buhler} introduce deep hedging, where a neural network hedging strategy is optimized under convex risk measures and proportional transaction costs. 
On these premises, a growing literature has focused on enhancing the methodological setting by employing state-of-the-art deep RL algorithms  \parencite{Du, caoDeepHedgingDerivatives2021,Cannelli}, %
which feature neural networks in the RL architecture, or by using risk-aware objectives \parencite{vittori_option_2020}.

Most studies mentioned above focus on simulation-based experiments, where synthetic data generated from arbitrage-free models is used both for training and testing purposes. Only a subset of the literature has also investigated RL-based hedging strategies in empirical settings, which is a key factor for the practical applicability of the approach.
An early example is given by \textcite{Buhler}, who hedge an at-the-money S\&P 500 call with the index alone, training on a volatility (GARCH) model fitted to past returns and testing on the subsequent historical path, where the deep hedge outperforms a BS hedge with filtered volatility.
\textcite{Giurca} show that a deep RL agent trained on synthetic data generated from an ensemble of diffusion models (Black-Scholes and Heston) successfully transfers the Delta hedging strategy to S\&P 500 options data. 
\textcite{xiao_optimal_2021} train on both synthetic and empirical S\&P 500 index option paths finding overall successful performance, but do not systematically compare results across the two training regimes. 
\textcite{xu_delta-gamma-like_2022} train and test exclusively on empirical data, covering S\&P 500, S\&P 100, and Dow Jones Industrial Average indices. Notably, the authors target Gamma risk alongside Delta, but they only rely on the underlying asset as hedging security, and include a Gamma-related penalty in the reward function.
\textcite{zernikovWhatDoesDeep2026} also employ a purely empirical training regime, using S\&P 500 daily data from 2015 to 2023. The author designs a walk-forward test, training on one year of data and testing on the following one, and identifies a systematic haircut in RL-hedging compared to BS Delta, as well as regime fragility across different years.
\textcite{mikkila23} provide the most systematic empirical study to date, explicitly comparing a simulation-trained and an empirically-trained agent on S\&P 500 intra-day data and finding superior performance for the latter.
While the result corroborates the potential of RL in tackling dynamic hedging in a pure data-driven way, we here advocate for a sim-to-real approach for two main reasons. First, real data are scarce and contingent. Synthetic environments, by contrast, accommodate the generation of arbitrarily many training episodes under arbitrary regimes, depending on the parametrization of the chosen model. Second, models afford control: before transferring the policy onto real data, one can gauge the algorithm's accuracy and reliability on the synthetic environment, which is fully understood and controllable.

While the empirical evaluation of RL policies is increasingly being addressed, the study of hedging strategies that use instruments beyond the underlying asset has received limited attention, and mostly in simulated environments. 
In the seminal paper that introduces deep hedging, \textcite{Buhler} validate the framework in a Heston setting where a variance swap is used as a second hedging instrument alongside the underlying. As anticipated above, the authors also test the framework on historical S\&P 500 index data, but they rely on a single-instrument Delta hedge for the empirical evaluation.  
Within the deep hedging framework, \textcite{armstrongDeepGammaHedging2024a} hedge a call option with the underlying and a second call, under uncertainty on the drift and volatility of the underlying. In a Black-Scholes simulated setting, the learned strategy approaches Gamma hedging when the loss rewards robustness to model uncertainty.
\textcite{cao_gamma_2023} use RL to target Gamma and Vega risk with an at-the-money hedging option. The agent decides only the option trade, while the Delta position is rebalanced exactly and at no cost at each step. Training and testing data are both generated under a SABR-type stochastic volatility model.
Closer to our empirical application, \textcite{francoisDeepHedgingOptions2025} test a deep RL agent on S\&P 500 index option data, hedging a straddle with the underlying and a short-dated option, and target Gamma risk against a Delta-Gamma benchmark. The authors train under the physical measure on an econometric model of the implied volatility surface and condition the agent on the fitted model factors. While the study shows that implied-volatility information improves performance, the state relies on model-specific quantities, and, since the simulator is not risk-neutral, one must check whether the agent harvests risk premia.
In contrast, we target Vega risk with a long-dated hedging option, restrict the agent to observable market quantities only, and train under calibrated risk-neutral dynamics, where the agent can improve its objective only by reducing variance and costs. %

A different route to data-driven hedging learns the market dynamics rather than the hedging policy. A few studies in this strand focus on hedging with options.
\textcite{cohenHedgingOptionBooks2022} fit neural-SDE market models to option prices and derive hedge ratios from the fitted model. On market data, the resulting hedges yield smaller errors than BS Delta and Delta-Vega hedging, and perform comparably to hedges based on the Heston model.
\textcite{contDatadrivenHedgingGenerative2025} simulate joint scenarios of the underlying and the implied volatility surface with a conditional generative model, and compute the hedge ratios that minimize risk across scenarios under transaction costs. Out of sample, on SPX options data, the authors report that the method matches and improves over Delta and Delta-Vega hedging.
Both studies compare against Delta-Vega hedging on market data, as we do, but compute the hedge at each date from a fitted model, rather than learning a hedging policy.

Outside reinforcement learning, our work relates to two classical literatures in mathematical finance, on hedging under transaction costs and on hedging volatility risk.
In the first, originated by \textcite{leland_option_1985}, the Black-Scholes hedge is adapted to discrete trading and proportional transaction costs, typically by restricting rebalancing to a no-trade region around the frictionless hedge \parencite{whalleyAsymptoticAnalysisOptimal1997a}.
Results of this kind are available for the Delta hedge; \textcite{herrmannModelUncertaintyRecalibration2017} describe their extension to Delta-Vega hedging as a major challenge.

In the second literature strand, different models have been proposed to overcome the constant-volatility misspecification of the BS framework. \textcite{crepeyDeltahedgingVegaRisk2004} groups them into three classes: the BS model used with each option's implied volatility, as in practitioner hedging (Section \ref{sec:pract_hedge}); local volatility models \parencite{dupirePricingSmile1994}; and stochastic volatility models, such as \textcite{Heston}, which we use in our simulation experiments. Among these, only stochastic volatility models account for volatility risk consistently, at the cost of complexity and market incompleteness \parencite{crepeyDeltahedgingVegaRisk2004}, which, as mentioned, can be addressed by introducing a plain-vanilla option as a tradeable instrument \parencite{romanoContingentClaimsMarket1997,davisMarketCompletionUsing2008}.
Using the underlying asset alone, only the part of volatility risk that is correlated with the underlying asset can be hedged, typically by adjusting the BS Delta for the expected change in volatility conditional on the move of the asset price \parencite{hull_optimal_2017}. Empirically, the (negative) correlation between asset returns and volatility is shown to capture only part of the variance risk premium \parencite{carrVarianceRiskPremiums2009}, suggesting an independent volatility risk factor.

By leveraging deep reinforcement learning, we optimize the hedging policy directly from data, thus accommodating discrete trading, transaction costs, and volatility risk in a unified framework. 

\section{Reinforcement Learning: Background and Notation}\label{sec:RL}
This section introduces the reinforcement learning notation used in the rest of the paper and the Proximal Policy Optimization (PPO) algorithm we adopt for optimizing the hedging policy. We refer to \textcite{Sutton} and \textcite{albrechtMultiAgentReinforcementLearning2024} for a complete treatment.

Reinforcement learning (RL) trains a computer model, referred to as an \textit{agent}, to choose a sequence of actions in an \textit{environment} so as to maximize a cumulative reward signal.
The RL problem is formalized as a discrete-time Markov Decision Process (MDP). Since hedging episodes have a finite number of decision dates $n$, we consider a finite horizon.
An MDP is a tuple $\langle \mathcal{S}, \mathcal{A}, \mathcal{P}, \mathcal{R}, \mathcal{P}_0 \rangle$, where $\mathcal{S}$ and $\mathcal{A}$ are the state and action spaces, $\mathcal{P}(s' \mid s, a)$ is the probability of moving to state $s_{i+1}=s' \in \mathcal{S}$ when action $a_i=a \in \mathcal{A}$ is taken in state $s_i=s \in \mathcal{S}$, $\mathcal{R}: \mathcal{S}\times\mathcal{A}\times\mathcal{S} \rightarrow \mathbb{R}$ is the reward function, and $\mathcal{P}_0$ is the distribution of the initial state $s_0$.
We write $r_{i+1} := \mathcal{R}(s_i, a_i, s_{i+1})$ for the reward collected after the transition from $s_i$ to $s_{i+1}$. This timing fits hedging, where the profit and loss of the position chosen at time $t_i$ is only known at $t_{i+1}$.
The agent does not know $\mathcal{P}$, and learns from sampled transitions only. This is the main difference between RL and dynamic programming, which solves the fully specified MDP directly.

The agent's behavior is described by a stochastic \textit{policy} $\pi(a \mid s)$, i.e.\ a probability distribution over actions $a \in \mathcal{A}$ conditional on the current state $s$. We write $\mathbb{E}_\pi[\cdot]$ for the expectation over trajectories $(s_0, a_0, s_1, \ldots, a_{n-1}, s_n)$ generated with $s_0 \sim \mathcal{P}_0$, $a_i \sim \pi(\cdot \mid s_i)$ and $s_{i+1} \sim \mathcal{P}(\cdot \mid s_i, a_i)$.
Given a discount factor $\gamma \in [0,1]$, the discounted return from time $t_i$ onward is
\begin{equation}
    G_i := \sum_{j=i}^{n-1} \gamma^{j-i} r_{j+1},
    \label{eq:return}
\end{equation}
and the agent seeks the policy that maximizes the expected discounted return of the episode
\begin{equation}
    J(\pi) := \mathbb{E}_\pi \left[ G_0 \right] = \mathbb{E}_\pi \left[ \sum_{i=0}^{n-1} \gamma^i r_{i+1} \right].
    \label{eq:RL_objective}
\end{equation}
The \textit{value} function and the \textit{action-value} function of $\pi$ are the expected return from time $t_i$ onward, conditional on the current state and on the current state-action pair, respectively,
\begin{equation}
    V_\pi(s) := \mathbb{E}_\pi \left[ G_i \mid s_i = s \right],
\end{equation}
\begin{equation}
    Q_\pi(s,a) := \mathbb{E}_\pi \left[ G_i \mid s_i = s,\, a_i = a \right].
\end{equation}
Both functions depend on the time $t_i$ only through the state, which includes the time to maturity (see Section \ref{sec:embed_DV_RL}).
Their difference
\begin{equation}
    A_\pi(s,a) := Q_\pi(s,a) - V_\pi(s)
\end{equation}
is the \textit{advantage} function, which indicates how much better or worse action $a$ is expected to perform relative to the average action drawn from $\pi$ in state $s$.

\paragraph{Policy gradient methods}
Value-based methods, such as the Q-learning approach of \textcite{Halperin} and the SARSA algorithm of \textcite{Kolm}, recover the policy from an estimate of the action-value function, by maximizing it over $\mathcal{A}$ at each state. This becomes impractical when actions are continuous, which is the case for hedge ratios.
Policy gradient methods instead parametrize the policy directly as $\pi_\theta$, with parameters $\theta \in \Theta$, and ascend the gradient of $J(\pi_\theta)$. By the Policy Gradient Theorem \parencite{suttonPolicyGradientMethods1999a}, this gradient can be written as\footnote{The theorem is usually stated with $Q_{\pi_\theta}$ in place of $A_{\pi_\theta}$. Subtracting the baseline $V_{\pi_\theta}(s)$ leaves the gradient unchanged, since $\mathbb{E}_{a \sim \pi_\theta(\cdot\mid s)}[\nabla_\theta \log \pi_\theta(a \mid s)] = 0$ for every $s$, and reduces the variance of its sample estimates \parencite{Sutton}. The relation holds up to a constant because the expectation averages over the steps of an episode. For $\gamma = 1$, the constant is the episode length $n$.}
\begin{equation}
    \nabla_\theta J(\pi_\theta) \propto \mathbb{E}_{\pi_\theta} \left[ A_{\pi_\theta}(s_i,a_i) \, \nabla_\theta \log \pi_\theta(a_i \mid s_i)  \right],
    \label{eq:policy_gradient}
\end{equation}
where the expectation is taken over the state-action pairs visited under $\pi_\theta$.
Equation \eqref{eq:policy_gradient} raises the probability of actions with positive advantage and lowers the probability of actions with negative advantage.

\paragraph{Actor-critic architecture}
Since $\mathcal{P}$ is unknown, the advantage has to be estimated from sampled transitions. Moreover, a tabular representation, storing one value per state, is not feasible when $\mathcal{S}$ is continuous, and would not generalize to states not visited during training. \textit{Deep} RL addresses this by parametrizing the relevant functions with neural networks.
In an \textit{actor-critic} architecture, the \textit{actor} network parametrizes the policy $\pi_\theta$, while the \textit{critic} network $V_\phi$, with parameters $\phi \in \Phi$, approximates the value function $V_{\pi_\theta}$. The critic enters the advantage estimate through the temporal-difference errors $\delta_i := r_{i+1} + \gamma V_\phi(s_{i+1}) - V_\phi(s_i)$, with $V_\phi(s_n) := 0$. Following the Generalized Advantage Estimation (GAE) approach \parencite{schulmanHighDimensionalContinuousControl2018}, we set
\begin{equation}
    \hat A_i := \sum_{j=i}^{n-1} (\gamma \lambda_{\text{GAE}})^{j-i} \, \delta_j.
    \label{eq:gae}
\end{equation}
The parameter $\lambda_{\text{GAE}} \in [0,1]$ trades off bias against variance, as $\lambda_{\text{GAE}} = 0$ gives the one-step estimate $\delta_i$, which relies entirely on the critic, while $\lambda_{\text{GAE}} = 1$ gives $G_i - V_\phi(s_i)$, which does not rely on the critic beyond the baseline.

\paragraph{Proximal Policy Optimization}
Policy gradient methods update $\theta$ using a local, first-order approximation of $J(\pi_\theta)$. Since $\pi_\theta$ can depend on $\theta$ in a highly nonlinear way, a small step in parameter space may induce a large change in the resulting policy. As the data are collected under the current policy, a bad update also degrades the data used for the next one, and cannot easily be corrected.
PPO \parencite{ppo} keeps each update close to the policy that collected the data, following the trust-region idea of \textcite{schulmanTrustRegionPolicy2017}, but replaces the constrained optimization with an unconstrained one that standard gradient-based optimizers solve directly.
Let $\pi_{\theta_{\text{old}}}$ denote the policy at the start of the current training iteration, and $\mathcal{D} := \{ (s^k_i, a^k_i, r^k_{i+1}, s^k_{i+1}) : i = 0, \ldots, n-1,\ k = 1, \ldots, N_{\text{batch}} \}$ the batch of transitions from $N_{\text{batch}}$ episodes collected under it, with $k$ indexing the episode. The parameters are then updated by $N_{\text{epochs}}$ passes of gradient ascent over $\mathcal{D}$. Define the probability ratio
\begin{equation}
\label{eq:ppo_ratio}
\omega_i(\theta) := \frac{\pi_\theta(a_i \mid s_i)}{\pi_{\theta_{\text{old}}}(a_i \mid s_i)}.
\end{equation}
The policy enters through the clipped objective
\begin{equation}
    \label{eq:ppo_clip}
    \mathcal{L}^{\text{CLIP}}(\theta) := \hat{\mathbb{E}}_{\mathcal{D}} \left[\min\Bigl(\omega_i(\theta)\, \hat A_i,\; \mathrm{clip}\bigl(\omega_i(\theta), 1-\varepsilon, 1+\varepsilon\bigr)\, \hat A_i \Bigr) \right],
\end{equation}
where $\hat{\mathbb{E}}_{\mathcal{D}}$ denotes the empirical average over the transitions in $\mathcal{D}$, $\hat A_i$ is the GAE estimate in Equation \eqref{eq:gae} computed under $\pi_{\theta_{\text{old}}}$, and $\varepsilon > 0$ controls the size of admissible policy updates.
Clipping $\omega_i(\theta)$ removes the incentive to push the ratio outside $[1-\varepsilon, 1+\varepsilon]$: once an update moves the policy too far from $\pi_{\theta_{\text{old}}}$, it stops improving the objective. Since the minimum is taken with the unclipped term, $\mathcal{L}^{\text{CLIP}}$ is a lower bound of the unclipped objective: changes in the ratio that would worsen the objective are never clipped.

Finally, actor and critic are trained jointly by maximizing
\begin{equation}
\label{eq:ppo_loss}
    \mathcal{L}(\theta, \phi) := \mathcal{L}^{\text{CLIP}}(\theta) - c_V \, \mathcal{L}^V(\phi), 
\end{equation}
where $\mathcal{L}^V(\phi) := \hat{\mathbb{E}}_{\mathcal{D}} \left[ \bigl( V_\phi(s_i) - G_i \bigr)^2 \right]$ is the critic loss, with return target $G_i$ from Equation \eqref{eq:return}, and $c_V > 0$ is its weight. \textcite{ppo} also add an entropy bonus that encourages exploration, which we do not use.

PPO has become a standard choice in continuous-control problems across various domains. In the option hedging literature, \textcite{Du} and \textcite{neaguDeepReinforcementLearning2025} benchmark several deep RL algorithms and rank PPO first and second best, respectively. For this reason, we adopt it in our Delta-Vega hedging framework.

\section{Option Hedging Setup}
\label{sec:setup}

We present our option hedging setting incrementally, with an eye toward  our reinforcement learning formulation of a two-instrument hedging strategy. 
Starting from the Black-Scholes framework, we progressively relax its key assumptions.
First, we account for non-constant volatility by adopting a stochastic volatility model, which in turn calls for a second hedging security.
Second, we incorporate discrete rebalancing and transaction costs, which render perfect replication unattainable and motivate the use of reinforcement learning as the solution framework.

\subsection{Delta-Vega Hedging}
\label{sec:delta_vega_hedging}

The Black-Scholes framework assumes the underlying asset dynamics are governed by a geometric Brownian motion,
\begin{equation}
\label{eq:GBM}
    dS_t = \mu S_t\,dt + \sigma S_t\,dW_t,
\end{equation}
where $\mu$ and $\sigma > 0$ are the constant drift and volatility, and $W_t$ is a standard Brownian motion under the physical probability measure $\mathbb{P}$. Under the risk-neutral measure $\mathbb{Q}$, the price of a European call option with strike $K$ and expiration $T$, evaluated at time $t \leq T$, admits the closed-form expression
\begin{equation}
\label{eq:BS}
\begin{gathered}
     C_t(S_t,\,T-t,\,K) = S_t\,\Phi(d_{1,t})
                         - e^{-r( T-t)}K\,\Phi(d_{2,t}), \\
     d_{1,t} = \frac{1}{\sigma\sqrt{T-t}}
              \left(\ln\frac{S_t}{K}
                    + \Bigl(r + \frac{\sigma^2}{2}\Bigr)(T-t)\right),
    \\
    d_{2,t} = d_{1,t} - \sigma\sqrt{T-t},
\end{gathered}
\end{equation}
where $\Phi(\cdot)$ is the standard normal cumulative distribution function and $r\geq0$ is the risk-free rate.\footnote{Without loss of generality, we will assume $r = 0$ throughout the rest of the paper.} 
Now consider a trader holding a short position in a call option. Over a discrete time interval 
$(t-\delta t, t]$, the option leg generates a profit-and-loss (P\&L) of $C_{t-\delta t} - C_t$. To offset this exposure, the trader holds over the interval a position $\mathcal{H}_t^S$ in the underlying asset, yielding a hedged portfolio P\&L of
\begin{equation}
\label{eq:simple_pnl}
    \mathrm{P\&L}_t = (C_{t-\delta t} - C_t) + \mathcal{H}_t^S(S_t - S_{t-\delta t}).
\end{equation}
In the continuous-time limit $\delta t \to 0$, the BS model admits a perfect hedge, since setting
\begin{equation}
\label{eq:delta}
    \mathcal{H}_t^S = \frac{\partial C_t}{\partial S_t} =: \Delta_t
\end{equation}
renders $\mathrm{P\&L}_t = 0$ to first order. The quantity $\Delta_t$ is the option's Delta.

The perfect hedge above relies on the assumption that the volatility $\sigma$ is constant (or, at least, deterministic in time). 
In reality, implied volatilities extracted from market option prices by inverting the BS formula vary across strikes and maturities, giving rise to the well-documented volatility smile and term structure, as well as over time, as market conditions evolve. 
A hedging strategy concerned with volatility risk must therefore be trained under a model that accounts for changes in volatility.

Specifically, we model the risk-neutral dynamics of the underlying asset with a stochastic volatility model, in which the variance itself is a stochastic process.
We adopt the seminal specification of \textcite{Heston}. Define $X_t := \ln S_t$, and let $\nu_t$ denote the instantaneous variance of the spot price at $t$, the SDE system reads
\begin{align}
    dX_t   & = - \frac{1}{2}\nu_t dt +  \sqrt{\nu_t} dW_t^S, \quad X_0 = 0, \ \label{eq:heston_dX} \\
    d\nu_t & = \kappa(\bar\nu - \nu_t)dt + \eta\sqrt{\nu_t}dW_t^\nu, \quad \nu_0 \geq 0, \label{eq:heston_dv}
\end{align}
where $W_t^S, W_t^\nu$ are two correlated Brownian motions under the risk-neutral probability measure $\mathbb{Q}$ with correlation $\rho \in [-1,1]$, and $\kappa, \bar \nu, \eta >0$.

When price dynamics are as in Equations \eqref{eq:heston_dX} and \eqref{eq:heston_dv}, the Brownian motion of the spot price is not the only source of stochasticity. To complete the market, i.e.,\ to perfectly hedge the option, the volatility risk (the Brownian increment $dW_t^\nu$ in \eqref{eq:heston_dv}) has to be accounted for as well. To this end, the hedging portfolio must include the underlying asset and another option (hereafter referred to as the hedging option) written on the same underlying, distinct from the target option\footnote{See \cite{romanoContingentClaimsMarket1997,davisMarketCompletionUsing2008} for theoretical results.}.
Let $\tilde C_t$ denote the price of the second option, and call $\mathcal{H}^{\tilde C}_{t}$ the position in the derivative at time $t$. Following a replicative portfolio argument, the portfolio holdings for $\tilde C_t$ and $S_t$ that yield a perfect hedge at each time $t$ are given by
\begin{align}
    \label{eq:heston_vega_hedging}
    \mathcal{H}^{\tilde C}_{t} & = \frac{\mathcal{V}_t}{\tilde{\mathcal V}_t}, \\
    \label{eq:heston_delta_hedging}
    \mathcal{H}^S_{t}  & = \Delta_t - \mathcal{H}^{\tilde C}_{t}\,\tilde\Delta_t,
\end{align}
where $\Delta_t$  is now the Heston Delta of the target option, and 
\begin{equation}
    \mathcal{V}_t := \frac{\partial C_t }{\partial \nu_t} \quad \quad \quad 
\end{equation} 
is the option's Vega ($\tilde\Delta_t, \tilde{\mathcal{V}}_t$ are the corresponding sensitivities of the hedging option). The joint rebalancing of the underlying asset and of the hedging option to cover for both the price and volatility risk is referred to as \textit{Delta-Vega} hedging\footnote{The hedge based on the sensitivity to the instantaneous variance (or spot volatility) is sometimes called the Delta-Sigma hedge, to distinguish it from Delta-Vega hedging, which uses the sensitivity to implied volatility \parencite{herrmannModelUncertaintyRecalibration2017}. For ease of notation, we will use the term Vega interchangeably, as also common in the literature \parencite{hullHedgingRisksWriting1987}, and we clarify the relationship between the two in Section \ref{sec:pract_hedge}.}. 
Under the Heston model, the option price and its partial derivatives are available in semi-analytical form, obtained by inverting the model's characteristic function. We therefore use this hedge as the benchmark against which we compare PPO in our in-model synthetic experiments \footnote{Needless to say, for a call there is a (static) perfect hedge given by the call-put parity relation. This, however, requires the availability of an option with equal contract specifics.}.

\subsection{Practitioner Hedging}
\label{sec:pract_hedge}
Equations \eqref{eq:heston_vega_hedging} and \eqref{eq:heston_delta_hedging} describe the optimal hedge under the Heston model, but require knowledge of the instantaneous variance $\nu_t$, whose process is not directly observable.
In practice, traders typically rely on the so-called \textit{practitioner} Black-Scholes hedging strategy, which treats the market option price as the output of the BS formula \eqref{eq:BS} evaluated at the market spot price $S_t$ and at the implied volatility $\sigma_t^{\rm imp}(K/S_t,T-t)$ 
for the contract's strike $K$ and time-to-maturity $T-t$. 
The sensitivities needed for hedging are thus available in closed-form,
\begin{align}
\label{eq:practitioner_greeks}
     \Delta^{\rm BS}_t & = \left. \frac{\partial C_t}{\partial S_t} \right|_{\sigma = \sigma_t^{\rm imp}} = \Phi(d_{1,t}), \\
     \mathcal{V}^{\rm BS}_t &  = \left. \frac{\partial C_t}{\partial \sigma} \right|_{\sigma = \sigma_t^{\rm imp}} = S_t \sqrt{T-t}\varphi(d_{1,t}),
\end{align}
where $\varphi(\cdot) = \Phi'(\cdot)$ is the standard normal density function.
Although not based on an internally consistent model, the practitioner strategy has the advantage of relying only on market-observable quantities, and it is the standard approach among option trading desks \parencite{hull_optimal_2017}. For this reason, we adopt it as the benchmark for our sim-to-real evaluation exercise where we test PPO on S\&P 500 market data. 
As in the Gamma-Vega study of \textcite{cao_gamma_2023}, as well as in most related RL works, our benchmarks do not account for transaction costs. Cost-aware versions of Delta-Vega hedging, such as no-trade regions around the frictionless hedge available for Delta hedging \parencite{whalleyAsymptoticAnalysisOptimal1997a}, have been described as a major open challenge \parencite{herrmannModelUncertaintyRecalibration2017}.

Since the agent is trained on data generated under the Heston model, it is useful to relate the Heston and the practitioner hedge.
For a fixed strike and maturity, the Heston price is the BS price evaluated at the model-implied volatility, $C_t = C^{\rm BS}(S_t,\sigma^{\rm imp}_t)$, where $\sigma^{\rm imp}_t$ is a function of $S_t$ and $\nu_t$. The chain rule gives
\begin{equation}
\label{eq:BS_Vega}
    \mathcal{V}_t = \mathcal{V}^{\rm BS}_t \, \frac{\partial \sigma^{\rm imp}_t}{\partial \nu_t},
\end{equation}
\begin{equation}
    \Delta_t = \Delta^{\rm BS}_t + \mathcal{V}^{\rm BS}_t \, \frac{\partial \sigma^{\rm imp}_t}{\partial S_t},
\end{equation}
and the hedge ratios in \eqref{eq:heston_vega_hedging}, \eqref{eq:heston_delta_hedging} become

\begin{equation}
    \label{eq:heston_pract_hedge_comparison}    
    \mathcal{H}^{\tilde C}_t  = \frac{\mathcal{V}^{\rm BS}_t}{\tilde{\mathcal{V}}^{\rm BS}_t} \, \chi^\nu_t, 
\end{equation}
\begin{equation}
    \label{eq:heston_pract_hedge_comparison_asset}
    \mathcal{H}^S_t  = \Delta^{\rm BS}_t - \mathcal{H}^{\tilde C}_t \tilde\Delta^{\rm BS}_t + \chi^S_t,
\end{equation}
with
\begin{equation}
    \chi^\nu_t = \frac{\partial \sigma^{\rm imp}_t / \partial \nu_t}{\partial \tilde\sigma^{\rm imp}_t / \partial \nu_t},
\end{equation}
\begin{equation}
    \chi^S_t =  \mathcal{V}^{\rm BS}_t \left(\frac{\partial \sigma^{\rm imp}_t}{\partial S_t} - \chi^\nu_t \, \frac{\partial \tilde\sigma^{\rm imp}_t}{\partial S_t}\right).
\end{equation}
The practitioner hedge corresponds to $\chi^S_t=0$ and $\chi^\nu_t=1$, that is, to implied volatilities that do not move with the spot (resulting in the so-called \textit{sticky-strike} volatility smile) and that move in parallel across maturities. 
Under Heston, the implied volatilities depend on both $S_t$ and $\nu_t$, with correcting factors $\chi^S_t$ and $\chi^\nu_t$ depending on the calibrated parameters and on the contract specifications.

\subsection{Hedging Under Transaction Costs}
Both the BS and Heston frameworks, despite their different treatment of volatility, share the assumptions of continuous-time trading and frictionless markets.
In reality, trades can occur only at discrete points in time, and portfolio managers face transaction costs whenever they change position in the market. 
Consequently, a trade-off emerges. On the one hand, perfect replication would require approaching the limit of continuous rebalancing by trading more frequently. On the other, continuous trading would make hedging infinitely costly.  The objective is to navigate this trade-off in a principled way --
a task for which reinforcement learning provides a natural
framework.

We consider an agent who holds a short position in an at-the-money (ATM) European call option with strike $K = S_0$ and expiration $T$. The agent enters the position at $t_0 = 0$ and dynamically rebalances at equispaced trading dates $0 = t_0 < t_1 < \ldots < t_n = T$\footnote{In the empirical application the hedging horizon need not coincide with the option's maturity, i.e., $t_n<T$. We address this in Section \ref{sec:res_simtoreal}.}. 
Throughout the rest of the paper, we denote $S_i := S_{t_i}$ for prices at date $t_i, \, i=0, \dots, n$, and $\tau_i := T-t_i$ for the time to maturity.
To hedge both price and volatility risk, the agent trades two instruments: the underlying asset and a hedging option written on the same underlying as the target call option. 
In line with standard practice, we assume the hedging option to be ATM (call) and have  maturity $\tilde{T} > T$, as longer-dated options are more sensitive to volatility risk (BS Vega \eqref{eq:BS_Vega} increases with the time-to-maturity) and are therefore more commonly used for Vega hedging \parencite{cao_gamma_2023}.

Let $\mathcal{H}_i^S$ and $\mathcal{H}_i^{\tilde{C}}$ denote the positions held over the interval $(t_{i-1}, t_i]$ in the underlying and hedging option, respectively, with $\mathcal{H}_1^S = h \in \mathbb{R} $ and $\mathcal{H}_1^{\tilde{C}} = \tilde h \in \mathbb{R}$.
The one-period portfolio profit-and-loss over $(t_{i-1}, t_i]$ is
\begin{equation}
\label{eq:reward_pnl}
\begin{aligned}
    \mathrm{P\&L}_i
    \;=\;&
    \underbrace{C_{i-1} - C_i}_{\text{option P\&L}}
    + \underbrace{\mathcal{H}_i^{S}(S_i - S_{i-1})}_{\text{asset P\&L}}
    + \underbrace{\mathcal{H}_i^{\tilde{C}}(\tilde{C}_i - \tilde{C}_{i-1})}_{\text{hedging option P\&L}} \\
    &\;
    - \underbrace{c\,\bigl|S_{i-1}\bigl(\mathcal{H}_i^{S} - \mathcal{H}_{i-1}^{S}\bigr)\bigr|}_{\text{asset costs}}
    - \underbrace{\tilde{c}\,\bigl|\tilde{C}_{i-1}\bigl(\mathcal{H}_i^{\tilde{C}} - \mathcal{H}_{i-1}^{\tilde{C}}\bigr)\bigr|}_{\text{hedging option costs}},
\end{aligned}
\end{equation}
where the parameters $c$ and $\tilde c$ capture proportional transaction costs as a fraction of the underlying asset and the option value, respectively.
Trades are executed at the rebalancing date $t_{i-1}$, when the position $\mathcal{H}_i$ is established, so that costs are proportional to the prices prevailing on that date. Our approach is similar to \textcite{mikkila23}, who also model proportional costs relative to the traded asset price, whilst \textcite{Kolm} consider a fixed tick-size cost per unit traded (which are comparable once $c \, S_{i} \approx \text{Tick Size}$).
Equivalently, $c$ can be interpreted as  the half bid-ask spread relative to the mid-price, so that $c = 1$ BPS represents the cost of crossing half a spread of one basis point of the instrument's mid-price. As commonly done in the related literature, we discard any explicit component of transaction costs such as commission and fees, and only model the spread, which typically dominates in liquid index products.

Given the step profit-and-loss defined above, the terminal wealth can be written as
\begin{equation}
    \label{eq:terminal_wealth}
    w_n = w_0 + \sum_{i=1}^n \mathrm{P\&L}_i,
\end{equation}
where $w_0=0$ under the assumption of a self-financing strategy.
Following \textcite{Kolm}, we model the agent's preferences via quadratic utility over terminal wealth. The hedging problem is therefore cast as the mean--variance optimization
\begin{equation}
\label{eq:max_utility}
    \max_{\{\mathcal{H}_i^S,\,\mathcal{H}_i^{\tilde{C}}\}_{i=1}^n}
    \Bigl(\mathbb{E}[w_n] - \lambda\,\mathbb{V}[w_n]\Bigr),
\end{equation}
where $\lambda > 0$ is the risk-aversion coefficient governing the
cost--variance trade-off. The expectation $\mathbb{E}[\cdot]$ and the variance $\mathbb{V}[\cdot]$ are taken with respect to the distribution of underlying and option prices (either from the market model or the empirical data), together with the positions chosen by the strategy.

\subsection{Embedding Delta-Vega Hedging in an RL framework}
\label{sec:embed_DV_RL}

In the following, we define the state space, action space, and reward function to cast the problem in Equation \eqref{eq:max_utility} as a reinforcement learning problem, and solve it via the PPO algorithm.

\subsubsection*{State Space}
For each time $t_i, \, i=0, \dots, n$, the state vector is defined as
\begin{equation}
\label{eq:def_state_vector}
    s_i := \bigl(S_i, C_i, \tilde C_i, \mathcal{H}_i^{S}, \mathcal{H}_i^{\tilde C}, \tau_i, \tilde\tau_i \bigr) \in \mathcal{S},
\end{equation}
\begin{equation}
\label{eq:def_state_space}
    \mathcal{S} := \mathbb{R}_+^3 \times [h_l,h_u]^2 \times [0,T] \times [0,\tilde T],
\end{equation}
where $S_i$ denotes the underlying spot price, $C_i$ the price of the target call option to be hedged, and $\tilde C_i$ the price of the hedging call option; $\mathcal{H}_i^{S}$ and $\mathcal{H}_i^{\tilde C}$ denote the positions in the underlying asset and in the hedging option held over $(t_{i-1},t_i]$, respectively, with bounds $h_l, h_u \in \mathbb{R}, h_l < h_u$; $\tau_i = T-t_i$ and $\tilde\tau_i = \tilde T-t_i$ denote the time-to-maturity of the main option and of the hedging option, where $T,\tilde T \in \mathbb{R}_+$, $T<\tilde T$, are the respective expiration dates 
in year terms.

The state space definition in Equation \eqref{eq:def_state_space} is completely detached from any model specification, and only includes variables that are directly observable from the market.
In particular, we stress that no latent or model-specific volatility variable is included in the state space. 
Such a choice is in contrast with related works using the Heston model as market simulator. For example, \textcite{Giurca} include the instantaneous variance of the log return process in the set of observables when training on the synthetic Heston environment, and then rely on the VIX index as a proxy for volatility when deploying the policy to S\&P 500 market data.
To avoid making any assumption on the volatility proxy, we do not enlarge the state space with the instantaneous variance $\nu_t$. 
Besides, the two option prices, with their respective times to maturity, carry information about the at-the-money implied volatility and its term structure implicitly. Such information is found to be relevant in \textcite{francoisDeepHedgingOptions2025}, who include the fitted factors of the implied volatility model they use for training in the state space. Here, the agent infers the volatility information from market prices in a model-free way.

\subsubsection*{Action Space}
At each date $t_i$, $i=0,\ldots,n-1$, the agent selects the portfolio holdings
\begin{equation}
    a_i := \bigl(\mathcal{H}_{i+1}^S, \mathcal{H}_{i+1}^{\tilde C}\bigr) \in \mathcal{A},
\end{equation}
\begin{equation}    
    \mathcal{A} := [h_l,h_u]^2 \subseteq \mathbb{R}^2.
\end{equation}

These are the positions in the underlying asset and in the hedging option held over the interval $(t_i,t_{i+1}]$. Throughout, we assume that the trader's book consists of a single normalized unit of the target option, hence $\mathcal{H}_i^S$ and $\mathcal{H}_i^{\tilde C}$ can be interpreted as hedge ratios, i.e.,\ units of the underlying and of the hedging option held per unit of the target option. 
The bounds $h_l$ and $h_u$ restrict the ratios to economically sensible values, which also improves sample efficiency and convergence \parencite{cao_gamma_2023}. Since the agent holds a short position in a European call option, the position's Delta is $-\Delta_{i} \in [-1,0]$, so the Delta hedge in the underlying lies in $[0,1]$.
Likewise, since the hedging option shares the same strike while $\tilde T > T$, the practitioner hedge ratio $\mathcal{V}^{\rm BS}_{i}/\tilde{\mathcal V}^{\rm BS}_{i}$ lies in $[0,1]$ as well, as $\tilde{\mathcal V}^{\rm BS}_{i} > \mathcal{V}^{\rm BS}_{i} \geq 0$ for near-the-money options. In our numerical experiments we thus set $h_l = 0$ and $h_u = 1$. Empirically, we find that the bounds do not bind for the Heston model hedge under the calibration and maturities of our synthetic tests, whereas with the longer maturities and multiple calibrations of the sim-to-real study the model hedge ratio for the hedging option can exceed one (see Section \ref{sec:res_emptests}).

\subsubsection*{Reward Function}
\label{sec:reward}
To reduce the optimization problem \eqref{eq:max_utility} to a per-step RL objective, we follow \textcite{Kolm} and define the instantaneous reward as a linear-quadratic function of the one-period $\text{P\&L}_{i+1}$ in Equation \eqref{eq:reward_pnl}, i.e., 
\begin{equation}
\label{eq:reward}
    \mathcal{R}(s_i,a_i,s_{i+1}) := \mathrm{P\&L}_{i+1} - \lambda\,\mathrm{P\&L}_{i+1}^2.
\end{equation}
\textcite{ritterMachineLearningTrading2017} shows that maximizing the expected sum of such rewards approximately solves the mean-variance problem \eqref{eq:max_utility}, and \textcite{Kolm} apply this result to option hedging. The approximation holds when per-step P\&L increments are uncorrelated across periods, so that $\mathbb{V}[w_n]$ is the sum of the per-step variances, and when the expected per-step P\&L is small, so that $\mathbb{E}[\left( \mathrm{P\&L}_{i+1} \right)^2]$ is close to $\mathbb{V}[\mathrm{P\&L}_{i+1}]$. The quadratic penalty with coefficient $\lambda$ therefore controls the cost--variance trade-off directly through the reward signal.

\section{Results}
\label{sec:results}

We conduct three main numerical experiments to evaluate our PPO Delta-Vega-like approach. In Section \ref{sec:res_frictionless}, we validate PPO against the most robust benchmark we have at our disposal, namely the Delta-Vega hedging strategy based on semi-analytical partial derivatives of the Heston model in the absence of transaction costs. We shall consider also the Delta-only hedging strategy (where only the underlier is traded as hedging instrument) as an additional benchmark, as it is the most familiar hedging scheme among researchers and practitioners and, most importantly, it exposes the costs and the benefits of holding a second hedging instrument.

In Section \ref{sec:res_costs}, we introduce transaction costs and study the mean-variance optimization performance of the PPO agent against the same two benchmark strategies. We present a grid of experiments combining different risk aversion levels for three distinct transaction cost scenarios, breaking up the roles of each hedging security.

Lastly, in Section \ref{sec:res_simtoreal} we study the sim-to-real transfer of the policy trained on synthetic data onto S\&P 500 market data. In this setting, we employ as benchmarks the practitioner Delta and Delta-Vega hedging strategies, based on Black-Scholes Greeks provided by the data vendor.

\subsection{Frictionless Environment}
\label{sec:res_frictionless}

The first test is meant to validate PPO in a controlled environment. Namely, we consider a Heston-simulated market under no-cost assumption ($c=\tilde c=0$ in Equation \eqref{eq:reward_pnl}), for which we know that the Delta-Vega hedging strategy is indeed the optimal policy, net of time discretization.
The benchmark portfolio positions are given by Equation \eqref{eq:heston_vega_hedging} and \eqref{eq:heston_delta_hedging}, where the partial derivatives are evaluated algorithmically, i.e.,\ by applying automatic differentiation to the option price as computed from the Heston model's characteristic function.

In the following examples, we consider a single underlying asset (for instance, a stock) generated under Heston dynamics. We simulate all training and testing episodes using a single parametrization, $\kappa=1.03, \bar \nu = 0.08, \eta = 1.0, \nu_0 = 0.04, \rho=-0.65$.\footnote{
Just to avoid choosing an arbitrary parametrization, we pick parameters calibrated to historical implied volatility surface (Euro Stoxx 50 index options on April 30, 2025), but we stress that in the synthetic experiments we do not focus on the specific market regime given by the Heston simulation. This will be the focus of the empirical application in Section \ref{sec:res_simtoreal}.}
Time is discrete and the agent observes 5 prices per day, so that the dates are equispaced, $t_i=i\,\delta t, \, i=0, \dots, n$, with $\delta t=1/(5\cdot 252)$ years, and an episode has $n=300$ steps. 
The underlying price at initial time is $S_0=100$. The option to be hedged is an at-the-money (ATM) call, meaning $K=100$, with $T=60$ days maturity. The second, hedging option is an ATM call with $\tilde T = 120$ days maturity.
Each episode terminates at the main option's expiration. The episode length, roughly a financial year-quarter, is in line with other applications in the literature \parencite{cao_gamma_2023, vittori_option_2020} and is the same we are going to use also in the empirical investigation. At the beginning of each episode, the agent holdings 
are set equal to the ones of the Delta-Vega benchmark. We note that this has no impact in the absence of transaction costs, as the agent learns to move directly toward the optimal positions within a single step, but is relevant under transaction frictions, as a position jump between subsequent steps results in large trading costs.

We train over a $N_{\text{train}}=40{,}000$-episodes dataset on a single CPU, and test on a deliberately large set,  $N_{\text{test}}=10{,}000$, to get reliable statistics. 
In this specific setting with no transaction costs involved, we set the discount factor $\gamma=0$. With $c=\tilde c=0$ in Equation \eqref{eq:reward_pnl}, the position chosen at date $t_i$ enters $\text{P\&L}_{i+1}$ and appears in no other term of the reward sequence, so each action affects only the immediately following step.
The hedging problem therefore decouples into $n$ independent one-step problems, each a static mean-variance optimization over the next price increment. 
In all other settings involving transaction costs, we set $\gamma = 0.99$. Although the hedging problem has a finite horizon and would theoretically  be unaffected by discounting with $\gamma = 1$, we follow the original implementation of PPO \parencite{ppo} and set $\gamma$ slightly below one for numerical stability of the value function estimate.
Other PPO hyperparameters and training performance are commented in Appendix \ref{app:hyper_train}.

\begin{figure}[ht]
    \centering
\includegraphics[width=1\linewidth]{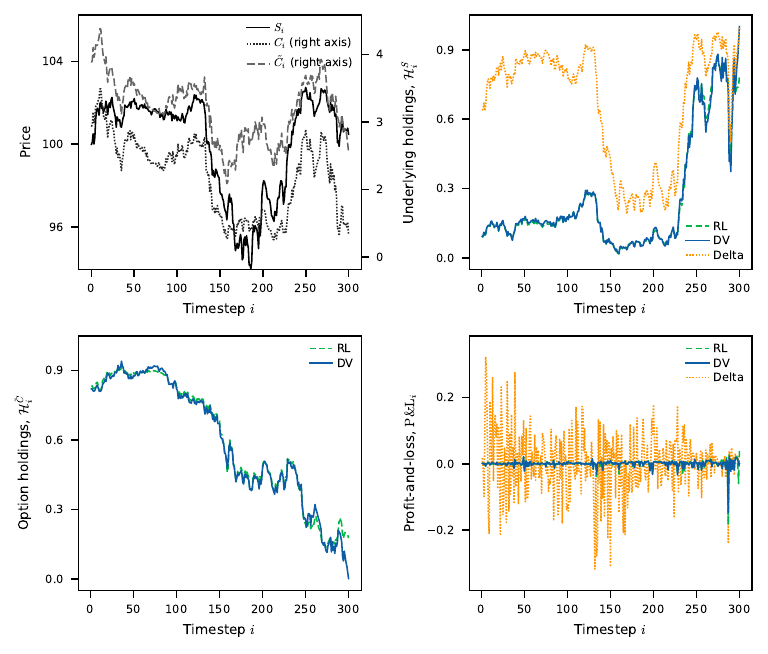}
    \caption{One out-of-sample episode from the frictionless test set, $c=\tilde c=0$. Top left panel: underlying asset price $S_i$ (left axis), main option price $C_i$, and second option price $\tilde C_i$ (right axis). Top right panel: hedging positions in the underlying asset, $\mathcal{H}_i^S$, for all hedging strategies: PPO (RL), Delta-Vega, and Delta-only. Bottom left: hedging positions in the second option $\mathcal{H}_i^{\tilde C}$ for PPO and Delta-Vega policies (Delta-only has zero exposure). Bottom right: step profit-and-loss $\text{P\&L}_i$ for all strategies.}
    \label{fig:traj_frictionless}
\end{figure}

Figure \ref{fig:traj_frictionless} displays a representative out-of-sample hedging episode. 
In the top left panel, we show the path of the market variables that enter the agent's observable space, namely the underlying asset price, the main option price (to be hedged), and the second option price (for hedging). 
In the top right panel, we compare the hedging position in the underlying asset selected by the agent at each time step, $a_i = \mathcal{H}_{i+1}^S$, with the one prescribed by the Delta-Vega hedging strategy and the Delta-only hedging strategy. Here, Delta-Vega is to be regarded as the optimal policy for the Heston stochastic volatility environment, net of discrete-time approximation error. As is evident, we find remarkable agreement between PPO's policy and the optimal one, whilst Delta hedging over-allocates to the underlying asset, in line with Equation \eqref{eq:heston_delta_hedging}. 
The bottom left panel displays the second policy dimension, namely the holdings in the hedging option, for both PPO and the optimal strategy (Delta-only has zero weight by definition). Again, PPO follows the optimal policy closely. 
We note that the deviation near the final steps of the episode is also observed in a fraction of other test episodes. Near expiration, the Delta of an ATM option becomes highly sensitive to small movements in the underlying (i.e.\, the option's Gamma is high), producing sharp changes in the optimal holdings. As these states are visited only once per episode, they are likely the most challenging region for the policy to learn accurately. 
Besides, even with the discount factor $\gamma=0$, RL optimizes the average reward signal from a batch of trajectories, while the semi-analytical benchmark is optimal point-wise, that is, at each step along every trajectory generated by the model.
Finally, the bottom right panel displays the profit-and-loss at each step of the episode for the three considered hedging policies (see Equation \eqref{eq:reward_pnl}), expressed as a percentage of the underlying asset price at time $t_0$. Although all strategies yield a zero-mean P\&L over time, Delta hedging clearly exhibits a noisier pattern, as expected in a stochastic volatility environment.

\begin{figure}[ht]
    \centering
    \includegraphics[width=1\linewidth]{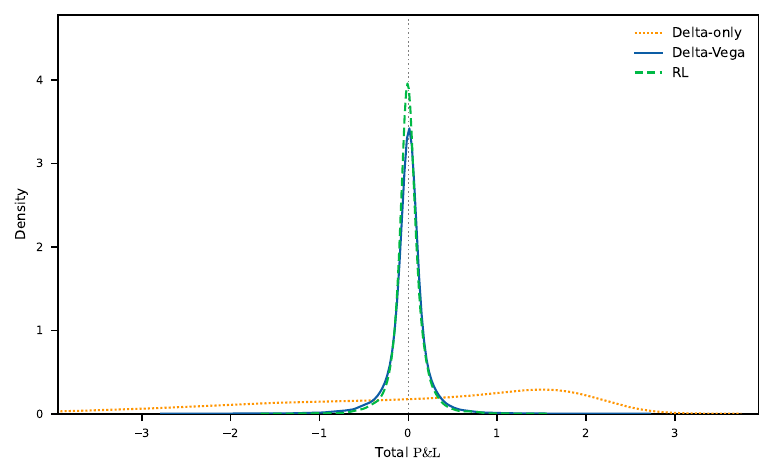}
    \caption{Total profit-and-loss 
    distribution for PPO (RL), Delta-only, and Delta-Vega hedging strategies in the frictionless Heston environment. 
    Curves are Kernel Density Estimates over a 10,000 episodes test set.}
    \label{fig:hist_frictionless}
\end{figure}

Figure \ref{fig:hist_frictionless} confirms our insights from the single-episode analysis above by displaying the distribution of each strategy's cumulative profit-and-loss $\sum_{i=1}^n\text{P\&L}_i$ over the $N_{\text{test}}=10,000$ test episodes. PPO distribution is in strong agreement with that of the semi-analytical Delta-Vega hedge, proving that the agent effectively replicates the optimal policy. 
Delta hedging, by contrast, exhibits markedly higher dispersion and pronounced negative skewness, which is consistent with the unhedged higher-order Gamma exposure of a short-call position.

\subsection{Introducing Costs}
\label{sec:res_costs}

In this section, we introduce transaction costs in the simulated market environment. 
As a baseline calibration, we set the cost rate on the underlying asset to $c = 1$ BPS and the option transaction cost to $\tilde{c} = 1\%$ in Equation \eqref{eq:reward_pnl}. This reflects a low-friction environment consistent with highly liquid index markets. The $1\%$ figure is of the same order as the half bid-ask spreads of SPX options documented in the empirical literature \parencite{kaeckPriceImpactBid2022}, and close to the quoted half-spread of the (hedging) options in our dataset (Section \ref{sec:data_preprocessing}).
As pointed out in \textcite{mikkila23}, the lower the costs, the more difficult for the RL agent to exploit cost-savings strategies (hence, the harder to outperform cost-unaware strategies).
The risk-aversion coefficient is set to $\lambda=1$. 

\begin{figure}[ht]
    \centering
    \includegraphics[width=\linewidth]{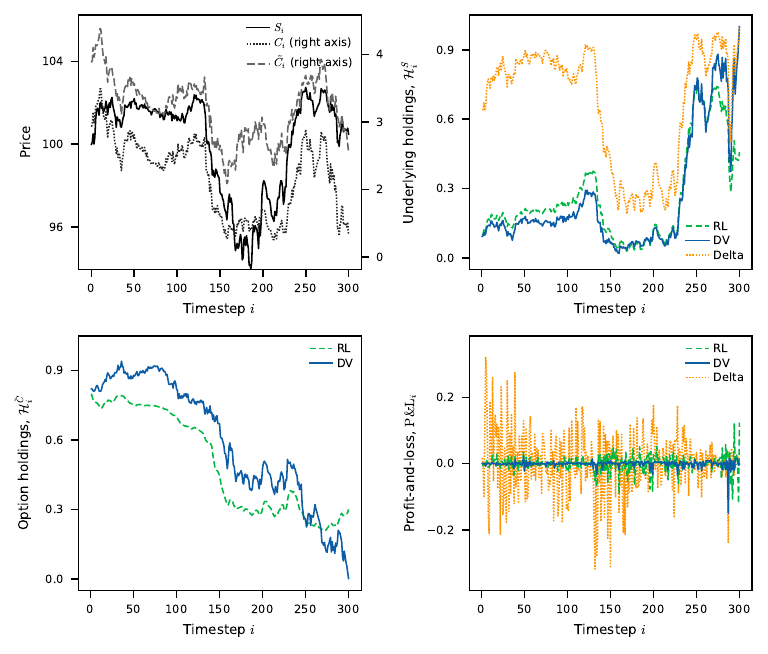}
      \caption{One out-of-sample episode under transaction costs ($c=1$BPS, $\tilde c=1\%$). Top left panel: underlying asset price $S_i$ (left axis), main option price $C_i$ (right axis), and second option price $\tilde C_i$ (right axis). Top right panel: hedging positions in the underlying asset $\mathcal{H}_i^S$ for all hedging strategies: PPO (RL), Delta-Vega, and Delta-only. Bottom left: hedging positions in the second option $\mathcal{H}_i^{\tilde C}$ for RL and Delta-Vega policies (Delta-only has zero exposure). Bottom right: step profit-and-loss  $\text{P\&L}_i$ for all strategies.}
    \label{fig:traj_cost}
\end{figure}

Figure \ref{fig:traj_cost} displays the very same market simulation used for the frictionless case in Figure \ref{fig:traj_frictionless} now in the presence of transaction costs. As evident from the top right and bottom left panels, which report the hedging positions in the underlying asset and option respectively, the PPO hedge no longer reproduces the Delta-Vega benchmark as closely as in the zero-costs scenario. This is expected, as Delta-Vega ceases to be the optimal policy once trading costs are introduced in the environment. In line with previous findings in the Delta-hedging related literature (e.g.,\ \cite{Kolm}), PPO seems to follow the benchmark positions in a more conservative way, trying to anticipate price movements and trading more smoothly to pay less transaction costs. In the specific example, we find this effect to be more evident in the option leg (bottom left panel) than for the underlying asset leg (top right), in line with option trading being relatively more expensive than trading in the asset (net of the different price scales of the two securities).
We point out that, in some regions of the state space, the agent even decides not to rebalance at all, as evident from the bottom left panel where PPO's option holdings remain roughly idle for some steps in the first half of the episode. 
By inspecting the step $\text{P\&L}_i$ path displayed at the bottom right panel, we can appreciate that the agent's deviation from the semi-analytical strategies may come at the expense of a more volatile profit-and-loss profile, especially close to expiration.

\begin{figure}[ht]
    \centering
    \includegraphics[width=1\linewidth]{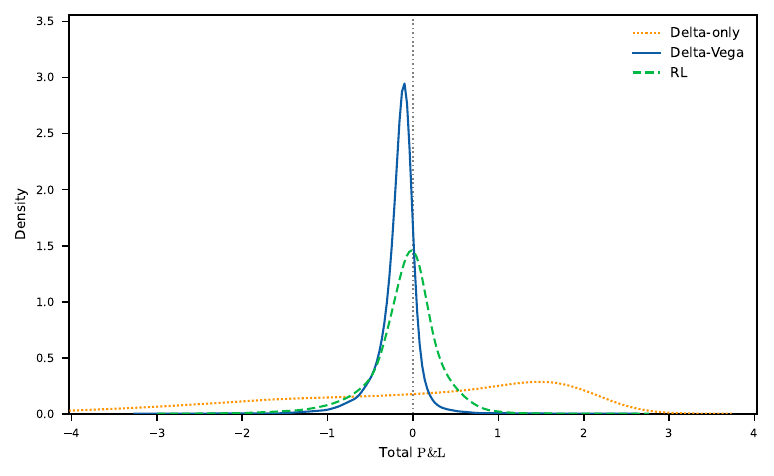}
    \caption{Total profit-and-loss distribution for PPO (RL), Delta-only, and Delta-Vega hedging strategies in the Heston environment with transaction costs $c=1$ BPS, $\tilde c=1\%$ (risk-aversion $\lambda=1.0$). 
    Curves are Kernel Density Estimates over a 10,000 episodes test set.}
    \label{fig:hist_costs}
\end{figure}

The trade-off between hedging costs and hedging error is made explicit in Figure \ref{fig:hist_costs}, which displays the cumulative P\&L distribution for each hedging strategy over the full test set. The Delta-Vega policy yields the lowest realized volatility in total P\&L among all strategies ($0.249\%$) but it is now characterized by a negative cumulative P\&L of $-0.170\%$, which is explained by the average cumulative transaction costs equal to $0.168\%$. PPO trades volatility for cost savings: the mean P\&L is $-0.085\%$ (achieving mean total costs of $0.079\%$ -- roughly half that of the benchmark), whilst standard deviation is now $0.396\%$. 

Below, we examine the mean-variance objective in more detail, extending the analysis across a grid of transaction cost and risk-aversion scenarios.

\subsubsection{Robustness to Transaction Costs}

To assess the robustness of the PPO agent under different market conditions,  we consider two additional transaction cost scenarios alongside the baseline calibration of $c = 1$ BPS and $\tilde{c} = 1\%$. In the first alternative scenario, we raise only the option cost $\tilde{c}$ from $1\%$ to $5\%$, keeping the asset cost at $c = 1$ BPS. This is representative of markets where the underlying is highly liquid but option bid-ask spreads are relatively wide, as observed in single-stock equity options \parencite{muravyevOptionsTradingCosts2020}.
In the second alternative scenario, we additionally raise the asset cost to $c = 5$ BPS. This level is consistent with the rescaled tick size of less liquid index futures such as the Euro Stoxx 50 \parencite{vittori_option_2020}. 
We deliberately exclude the high-asset-cost, low-option-cost scenario from our analysis, as it does not reflect the typical cost structure of option markets, where bid-ask spreads on options are generally wider than on the underlying. 

\begin{table}[ht]
\centering
\caption{%
Hedging performance across transaction-cost regimes on 10,000 episodes test set generated under the Heston model. Reported values are aggregated across four independent training runs, at risk-aversion $\lambda=1$. All quantities are in percent of $S_0$. Bold marks the best (highest) mean-variance objective in each cost scenario.}
\label{tab:inmodel_results_60d_spx_short_nonhp_agg_ra1}

\begin{tabular}{@{}l
    S[table-format=-1.3] S[table-format=1.3]
    S[table-format=1.3] S[table-format=1.2]
    S[table-format=1.3] S[table-format=1.2]
    S[table-format=-2.2]
  @{}}
\toprule
\multirow{2}{*}{Strategy} & \multicolumn{2}{c}{Total P\&L} & \multicolumn{2}{c}{Asset market} & \multicolumn{2}{c}{Option market} & \multicolumn{1}{c}{\multirow{2}{*}{Objective}} \\
\cmidrule(lr){2-3} \cmidrule(lr){4-5} \cmidrule(lr){6-7}
 & {Mean} & {Std} & {Cost} & {Turnover} & {Cost} & {Turnover} & \\
\midrule
\multicolumn{8}{@{}c}{\textit{Low transaction costs} ($c=1$ BPS, $\tilde c= 1\%$)} \\[1pt]
Delta      & -0.047 & 1.719 & 0.052 & 6.51 & 0.000 & 0.00 & -3.00 \\
Delta-Vega & -0.170 & 0.249 & 0.036 & 4.27 & 0.132 & 4.22 & $\bm{-0.23}$ \\
RL         & -0.080 & 0.429 & 0.025 & 3.07 & 0.055 & 2.33 & -0.26 \\
\midrule

\multicolumn{8}{@{}c}{\textit{Mild transaction costs} ($c=1$ BPS, $\tilde c= 5\%$)} \\[1pt]
Delta      & -0.047 & 1.719 & 0.052 & 6.51 & 0.000 & 0.00 & -3.00 \\
Delta-Vega & -0.697 & 0.419 & 0.036 & 4.27 & 0.659 & 4.22 & -0.87 \\
RL         & -0.277 & 0.482 & 0.024 & 2.91 & 0.251 & 2.35 & $\bm{-0.51}$ \\
\midrule

\multicolumn{8}{@{}c}{\textit{High transaction costs} ($c=5$ BPS, $\tilde c= 5\%$)} \\[1pt]
Delta      & -0.256 & 1.776 & 0.261 & 6.51 & 0.000 & 0.00 & -3.41 \\
Delta-Vega & -0.841 & 0.482 & 0.180 & 4.27 & 0.659 & 4.22 & -1.07 \\
RL         & -0.350 & 0.460 & 0.096 & 2.41 & 0.252 & 2.35 & $\bm{-0.56}$ \\
\bottomrule
\end{tabular}
\end{table}

Table \ref{tab:inmodel_results_60d_spx_short_nonhp_agg_ra1} compares the RL agent hedging performance with the two benchmark strategies in the three friction scenarios. The reported figures are the average of four independent training runs, each evaluated on the same $N_{\text{test}}=10{,}000$-episodes test set.
In addition to the mean total P\&L and its standard deviation, we also show the test-set average of total cost and total absolute position change per episode, defined as $\text{Turnover} = 
\sum_{i=1}^{n}\bigl|\mathcal{H}_{i}- \mathcal{H}_{i-1}\bigr|$. This second metric measures the cumulative trading activity over an episode independently of price fluctuations, and therefore complements total cost as a summary of the agent's rebalancing behavior. In particular, it allows us to quantify the degree to which the RL agent smooths its trading trajectory relative to the benchmarks (the behavior that we already identified from the visual analysis of Figure \ref{fig:traj_cost}).
Both cost and turnover metrics are reported separately for the underlying and the hedging option to isolate the contribution of each instrument.
The first block in the table corresponds to the low-cost setting discussed in Section \ref{sec:res_costs}, now averaged over four independent training runs. The mean-variance trade-off is here summarized in the final column, that reports the objective function in Equation \eqref{eq:max_utility} evaluated over the test data. In this scenario, the PPO agent achieves a comparable, albeit slightly lower objective value than the Delta-Vega strategy, while clearly outperforming the Delta-only benchmark. This is expected, as the Delta-Vega policy should theoretically remain close to optimal when transaction costs are low. In fact, PPO incurs lower costs in both markets compared to the benchmark strategies (Delta has zero option costs by definition), yet the absolute amount of savings is negligible under the current friction level. Both total cost and turnover suggest that the agent adopted a more conservative trading strategy in the option market than in the underlying asset one, as previously suggested by the visual inspection of the single out-of-sample episode above (see bottom left panel in Figure \ref{fig:traj_cost}).
In the second cost block, with higher option costs $    \tilde c=5\%$, PPO consistently beats both benchmark strategies. It is worth noting that, although the agent does achieve lower trading costs than the benchmark in the pricier option leg, its edge with respect to the Delta-Vega strategy is not solely cost-related. The P\&L volatility, that the agent maintains stable from the low-cost scenario level of $0.429\%$, to the mid-cost one of $0.482\%$, displays an evident escalation for Delta-Vega, which shifts from $0.249\%$ to $0.419\%$. Since costs enter the P\&L directly (Equation \eqref{eq:reward_pnl}), when the cost rate increases, the P\&L realized volatility of a strategy that does not adapt its trading rises too. On the other hand, PPO manages to achieve lower costs while maintaining realized volatility comparable to the Delta-Vega level, thus yielding a higher objective value.
When also the underlying cost rate increases (bottom block in the table), the same effect makes PPO outperform the Delta-Vega strategy both in terms of mean and volatility, resulting in the highest margin in the objective value among the three friction scenarios. 

Interestingly, PPO's turnover in the hedging option remains essentially flat throughout the three different cost settings, suggesting 
that additional factors govern this dimension of the agent's policy (we turn to this in Section \ref{sec:res_sentitivity_risk_aversion}).  Furthermore, the propensity to a smoother trading in the option leg relative to the asset that we previously identified is no longer evident once the asset cost rate grows: in the third scenario, the turnover difference between RL and Delta-Vega is comparable between the asset ($2.41$ vs. $4.27$) and the option leg ($2.35$ vs $4.22$).

\begin{figure}
    \centering \includegraphics[width=\linewidth]{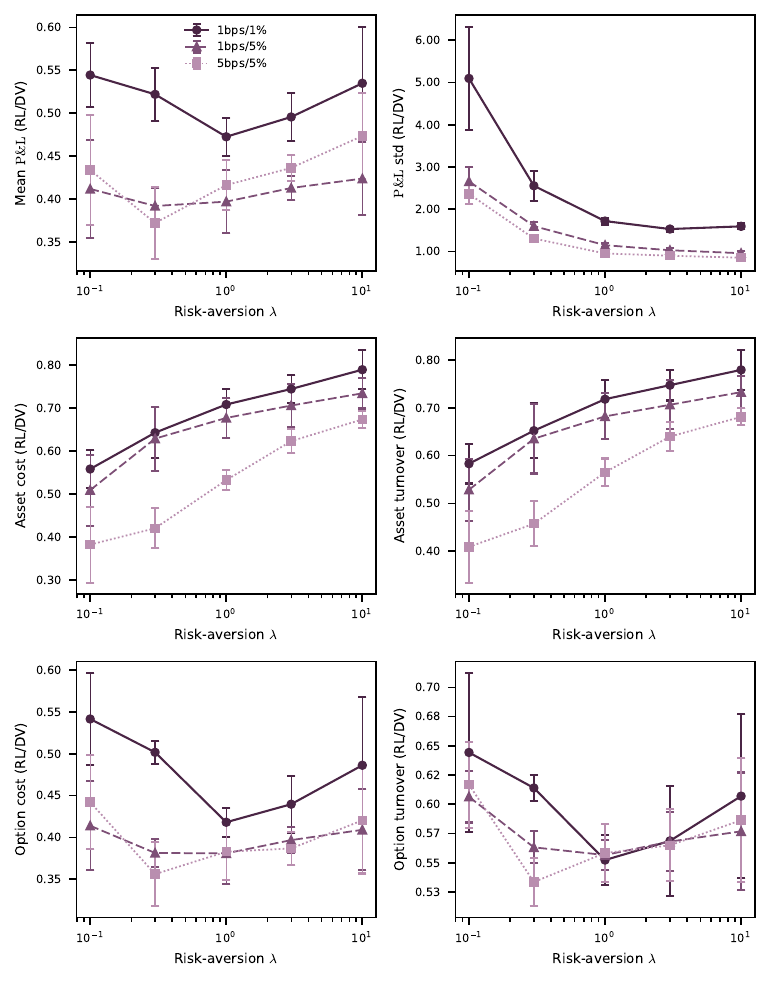}
      \caption{Effect of risk-aversion $\lambda$ on PPO (RL) out-of-sample hedging performance, relative to Heston's Delta-Vega strategy. Top row: mean P\&L (left) and P\&L volatility (right). Middle row: transaction cost (left) and turnover (right) on the asset leg. Bottom row: transaction cost (left) and turnover (right) on the hedging option leg. All quantities are shown across log-spaced risk-aversion levels $\lambda$ for the three cost scenarios (four independent runs each). Each curve is normalized by its own tier's Delta-Vega value, so $y=1$ corresponds to Delta-Vega in every panel, though at a different absolute level per tier (see Table \ref{tab:inmodel_results_60d_spx_short_nonhp_agg_ra1}). Bands show one standard deviation from the mean across runs.}
    \label{fig:ra_vs_costs}

\end{figure}

\subsubsection{Sensitivity to Risk Aversion}
\label{sec:res_sentitivity_risk_aversion}

As a final validation, we study the sensitivity of the agent's hedging performance to the risk-aversion parameter $\lambda$, which directly governs the cost--variance trade-off of Equation \eqref{eq:max_utility}.
Figure \ref{fig:ra_vs_costs} reports the hedging statistics of Table \ref{tab:inmodel_results_60d_spx_short_nonhp_agg_ra1} at different values of risk-aversion coefficients across the three friction scenarios. Each point is normalized by the corresponding scenario's Delta-Vega metric. 
In the top row, we plot the average total P\&L (left) and its standard deviation (right) across out-of-sample episodes. The volatility exhibits a clear monotonic relationship with the risk-aversion coefficient, decreasing as risk aversion increases. This is the expected behavior, since the agent is increasingly penalized for the quadratic term in the reward function as per Equation \eqref{eq:reward}. At low transaction costs, the agent's volatility plateaus above the Delta-Vega benchmark level. In the highest-friction scenario, instead, it matches the benchmark at $\lambda=3$ and falls slightly below it thereafter. As anticipated in the above analysis, this is driven by increasing volatility in the benchmark strategy, rather than solely PPO tightening. 
In the top-left panel, mean P\&L varies less systematically with the risk-aversion coefficient. Rather than a monotonic decline, we observe a turning point whose location shifts across cost scenarios.

The picture sharpens once we turn to transaction costs and turnover, shown in the middle and bottom rows for the asset and option leg respectively. The asset leg (middle row) tracks risk-aversion monotonically: RL's Delta-hedging cost and turnover both rise steadily with $\lambda$ across all three scenarios, mirroring the volatility panel's evidence of tighter tracking under stronger risk aversion. 
The option leg (bottom row) behaves differently. Cost and turnover first fall and then rise again, with the minimum at $\lambda=0.3$ in the high-friction scenario and at $\lambda=1$ in the low- and mid-friction ones.
Rather than a difficulty specific to learning the Vega hedge, this pattern is consistent with the cost asymmetry embedded in Equation \eqref{eq:reward_pnl}. Trading the hedging option is structurally more expensive than trading the underlying. Its proportional cost, $\tilde c \in \{1\%, 5\%\}$, is two orders of magnitude larger than that of the underlying, $c \in \{1\text{BPS}, 5\text{BPS}\}$, while its price is only about 20 times smaller ($\tilde C_0 \approx 4.7$ against $S_0 = 100$). Per unit traded, this leaves a net cost differential of roughly five times in the low-friction scenario.
The behavior of the agent fits this asymmetry. As $\lambda$ increases from its lowest values, Delta-hedging turnover rises sharply while Vega-hedging turnover falls, which indicates that the additional risk reduction is achieved mainly by tightening the cheaper asset leg. Once further tightening of the Delta hedge yields diminishing risk reduction, the agent starts to pay for additional Vega hedging, and option-leg cost and turnover rise at higher $\lambda$. The same mechanism accounts for the earlier turning point in the high-friction scenario. When trading the underlying is costlier (higher $c$), further tightening of the Delta hedge becomes expensive sooner, so the agent turns to the Vega leg at a lower $\lambda$.

Overall, PPO's transaction costs remain below Delta-Vega's throughout the whole $\lambda$ range and across all three scenarios, in both the asset (at most $79\%$ of the benchmark level) and option market (at most $55\%$). 
We do not report Delta-only values in this figure for the sake of visualization; however, we note that both PPO's asset-related costs and P\&L volatility remain below Delta-only's throughout the considered $\lambda$ range.

\subsection{Sim-to-Real Transfer}
\label{sec:res_simtoreal}

In the following, we train the PPO agent 
using data generated from the Heston stochastic volatility model and study the transfer of the policy to real SPX option market data. In Section \ref{sec:data_preprocessing} we describe the empirical dataset. In Section \ref{sec:emp_train} we outline how we build the synthetic training set specific for the empirical transfer task. Finally, Section \ref{sec:res_emptests} reports out-of-sample results.

\subsubsection{Data} 
\label{sec:data_preprocessing}

We use daily S\&P 500 index (SPX) option quotes covering trading days from January 2018 to July 2022 \parencite{historicaloptiondata}. For each contract we observe the underlying asset (index) level, strike, expiration date, option type, bid and ask prices, traded volume, open interest, and the corresponding Greeks (Delta and Vega) as computed by the data provider. All profit-and-loss figures reported in this paper are computed from quoted mid prices. 
As a preliminary step we discard quotes with a non-positive bid or ask, or with the ask below the bid, which we take as evidence of a misreported record. We hold out 2018 entirely for calibrating the Heston model used to generate the synthetic training data (discussed in Section \ref{sec:emp_train} below), and use 2019-2022 as an out-of-sample test set for the trained PPO agent.

\begin{table}[t]
\centering
\caption{SPX test sample (2019--2022) used to evaluate the PPO hedging policy.
$\tau_0, \tilde \tau_0$ denote the time-to-maturity of main option and hedging option at episode start in calendar days, respectively,  $\log | (K/S_0)|$ denotes absolute log-moneyness. Half-spread is relative, quoted half-spread of the hedging option $(\tilde C^\text{Ask}-\tilde C^\text{Bid})/(\tilde C^\text{Ask}+\tilde C^\text{Bid})$ averaged over all steps. Reported figures are mean values across episodes, with one standard deviation in parentheses below.}
\label{tab:spx_test}

\begin{tabular}{lcccccc}
\toprule
Year & Windows & Episodes & $\tau_0$ (days) & $\tilde \tau_0$ (days) & $|\log (K/S_0)|$ & Half-spread (\%) \\
\midrule
2019  & 4  & 188   & 263.3 & 437.7 & 0.015 & 0.56 \\
      &    &       & (126.6) & (173.9) & (0.009) & (0.19) \\[2pt]
2020  & 4  & 211   & 235.7 & 430.9 & 0.015 & 1.94 \\
      &    &       & (124.0) & (155.7) & (0.009) & (1.80) \\[2pt]
2021  & 4  & 445   & 272.7 & 450.5 & 0.015 & 0.96 \\
      &    &       & (117.7) & (147.3) & (0.009) & (1.00) \\[2pt]
2022  & 3  & 349   & 284.0 & 461.5 & 0.015 & 0.90 \\
      &    &       & (109.9) & (143.3) & (0.009) & (0.59) \\[2pt]
\midrule
Total & 15 & 1{,}193 & 268.0 & 448.2 & 0.015 & 1.05 \\
      &    &         & (119.2) & (152.6) & (0.009) & (1.11) \\
\bottomrule
\end{tabular}
\end{table}

We build hedging episodes by taking consecutive, non-overlapping windows of 60 trading days starting January 2019.
Time horizon is thus consistent with previous synthetic investigations; only the frequency is lower, from five to one step per day.
Within a window, a contract must carry a valid quote on every trading day to be eligible (an isolated gap of two days at most is bridged using the last available quote, while a contract with a longer gap is dropped from that window entirely -- 0.2\% of all daily observations in the final sample are bridged). For each eligible main contract we select a second contract used for Vega hedging. The second contract is required to expire at least 30 calendar days later than the main contract, so that the two legs carry meaningfully different sensitivity to volatility along the maturity dimension, and to share a similar moneyness with the main contract at the start of the episode (within two percentage points). When several contracts satisfy both conditions we favor, in order, the one closest in moneyness to the main contract, the one with higher open interest, and the one with the tighter bid-ask spread. 

Prices for the underlying, the main option, and the hedging option are all normalized by the index value at episode inception, $S_0$, and rescaled by a factor of 100, so that $S_0=100$ in every episode. This preserves the relative scale across the three instruments and lets P\&L be reported as a percentage of $S_0$ consistently across episodes with different index levels. Importantly, it also matches the convention used for the synthetic episodes in the training set, so that the actor and critic networks face inputs on the same scale in training and in testing. 

We apply two further filters, chosen to match the synthetic training data distribution and thus to prevent any substantial mismatch between train and test data. First, we restrict to episodes whose main contract is at the money at inception ($|\log (K/S_0)| \le 0.03$), matching the assumption (used throughout the previous sections) that the hedged option is ATM at the beginning of each episode. Second, we cap main and hedging option's time to maturity at inception to 600 and 800 calendar days, respectively, so that we are able to test the agent on a sufficiently wide, yet not excessive maturity range. 

Together, these filters yield 1{,}193 test episodes across 15 independent time windows. 
Clearly, episodes within a time window should not be regarded as independent draws, since they share the same history of the S\&P 500 index.
We acknowledge this limitation especially in the context of year-by-year analysis in Section \ref{sec:res_emptests}, where the number of independent windows is even smaller (about four per year).
Related works address this by considering longer time horizons \parencite{xu_delta-gamma-like_2022}, or by leveraging higher observation frequency \parencite{mikkila23} to account for more independent paths. \textcite{xiao_optimal_2021}, who also deal with daily observations, follow a similar approach as ours for building parallel episodes, pointing out that including the strike price $K$ in the agent's state space would allow including also out- and in-the-money contracts -- a route that we do not explore here. 

Table \ref{tab:spx_test}
reports summary statistics for the resulting sample by year.
Years 2021 and 2022 are the ones where most episodes in the dataset are concentrated. The statistics of the contract specifics, namely expiration and moneyness, are nevertheless comparable across years, due to the filtering applied above. The average difference between the main option and the hedging option expirations is of 180 calendar days, i.e.,\ roughly half a year, and moneyness is approximately centered around zero.
The aggregate half-spread, computed as $(\tilde C^\text{Ask}-\tilde C^\text{Bid})/(\tilde C^\text{Ask}+\tilde C^\text{Bid})$, averages to $1.05\%$, which corroborates our cost rate assumption of $\tilde c=1\%$. Spread data is aligned with market regimes across years: we record the lowest level ($0.56\%$) in 2019, when the stock market has been relatively stable, and the highest value ($1.94\%$) in 2020, when the market has been hit by the pandemic shock. 

\subsubsection{Synthetic Training Set}
\label{sec:emp_train}

We generate a training set specific to the sim-to-real transfer application. While still employing the Heston model as a market simulator as in previous sections, we now adapt the simulation setup to reproduce both the structure of the empirical dataset and the characteristics of its underlying asset.
We consider 60 days-long episodes at daily rebalancing frequency, where the option is ATM at $t_0=0$. Since contracts in the empirical set are not bound to expire by the end of each episode, and in fact are characterized by a variety of different expirations, we consider eight different maturity buckets, $\tau_0 \in [60,100,145,195,250,305,360,420]$ (trading days), to broadly match the empirical support shown in Table \ref{tab:spx_test}. The gap between the main option and second option maturities is assumed to be constant among the buckets and equal to 130 trading days, which again reflects the average difference  $\tilde \tau_0 - \tau_0$ (in trading days terms) observed in the data. We only consider one transaction cost scenario, namely $c=1$ BPS, $\tilde c=1\%$, which is consistent with the empirical estimate from SPX market data in Table \ref{tab:spx_test} above. We pick the risk-aversion coefficient $\lambda=0.5$, which is aligned to the typical values assumed by the first two P\&L moments in the environment at hand (mean per-step P\&L volatility for Delta Vega over the train set is $0.07\%$, while in the episodes with five steps per day it was $0.018\%$).
In this empirical section of the paper, we do not investigate the agent's sensitivity to risk-aversion levels.

\begin{table}[t]
\centering
\caption{Heston model parameters calibrated to 2018 daily SPX implied-volatility surfaces (one calibration per trading day, for 251 days). We report optimization bounds and summary statistics for each parameter.}
\label{tab:heston_calib}

\begin{tabular}{
    l
    S[table-format=-1.3]
    S[table-format=1.3]
    S[table-format=1.4]
    S[table-format=1.4]
    S[table-format=1.4]
    S[table-format=1.4]
}
\toprule
Parameter & {Lower bound} & {Upper bound} & {Mean} & {Std} & {Min} & {Max} \\
\midrule
$\kappa$ (mean reversion)
    & 1.00   & 8.00  & 4.10   & 1.52   & 1.00   & 7.37   \\
$\bar\nu$ (long-run variance)
    & 0.01  & 0.10 & 0.0470 & 0.0054 & 0.0376 & 0.0746 \\
$\eta$ (vol-of-vol)
    & 0.10  & 3.00 & 1.235  & 0.510  & 0.371  & 3.000  \\
$\nu_0$ (initial variance)
    & 0.001  & 0.10 & 0.0230 & 0.0200 & 0.0012 & 0.1000 \\
$\rho$ (correlation)
    & -0.90 & -0.10 & -0.806 & 0.055  & -0.900 & -0.628 \\
\bottomrule
\end{tabular}
\end{table}

To calibrate the Heston model used for generating synthetic training paths, we build a liquid daily implied-volatility surface for each of the 251 SPX trading days in 2018 from the same underlying quotes, keeping strikes with adequate liquidity (open interest of at least 500 contracts or volume of at least 10), a relative spread no wider than 15\%, and moneyness within a moderate band around the money, retaining the 25 most liquid strikes per expiry. Altogether, these filters result in approximately 220 usable quotes per day, on average. 
The five Heston parameters $(\kappa, \bar\nu, \eta, \nu_0, \rho)$ are then calibrated to each day's surface by minimizing squared implied-volatility error, using a neural-network surrogate for the Heston price in place of direct numerical integration, from five random starting points per day, keeping the best fit. All 251 days converge successfully. Table \ref{tab:heston_calib} summarizes the resulting distribution of daily parameters.

We generate $N_{\text{train}}=360{,}000$
episodes, with the training data allocated equally across the 251 calibrated parameter sets. The dataset size is augmented with respect to the one used in the previous runs ($N_{\text{train}}=40{,}000$) for two main reasons. First, episodes contain fewer steps than in the previous setting (from five steps per day to one): to ensure robust learning, the number of episodes per training batch is increased from $N_{\text{batch}}=128$ to $256$. Second, the learning objective is harder: training paths now originate from a set of different model parameters, making the inference of the non-observable instantaneous variance process even less trivial than before. Thus, we should also be able to train over a larger number of iterations.

In Appendix \ref{app:hyper_train} we report PPO hyperparameters and comment on training performance.

\subsubsection{Empirical Tests}
\label{sec:res_emptests}

\begin{figure}[ht]
    \centering
    \includegraphics[width=1\linewidth]{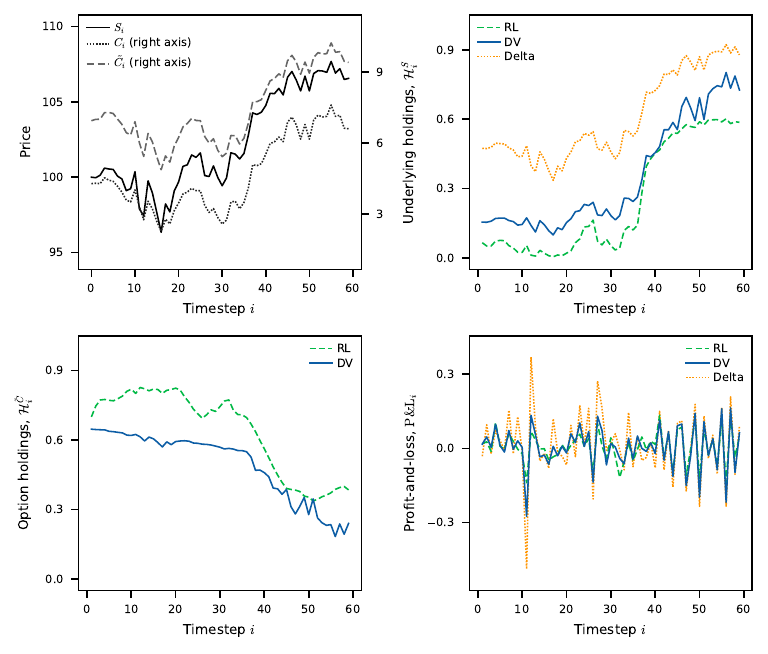}
    \caption{Out-of-sample hedging episode on S\&P 500 index data from February 9, 2021 to May 5, 2021. Main option (SPX210618C03940000): $\log(K/S_0)=0.007$, $\tau_0 = 129$ calendar days; hedging option (SPX211217C03950000): $\log(K/S_0)=0.01$, $\tilde \tau_0 = 311$. Transaction costs are assumed to be $c=1$ BPS, $\tilde c=1\%$. Risk-aversion is $\lambda=0.5$.}
    \label{fig:spx_traj}
\end{figure}

After having trained the PPO agent on Heston-simulated episodes, we test the policy out-of-sample on S\&P 500 market data. 
The benchmark policies are now the practitioner BS Delta and Delta-Vega hedging strategies described in Section \ref{sec:pract_hedge}.

As in previous sections, we begin by displaying a representative path from the test set in Figure \ref{fig:spx_traj}. 
The episode tracks the S\&P 500 index from February 9, 2021 to May 5, 2021, and reflects a stable market condition where the index increased by 6.5\%. The main contract (SPX210618C03940000), in which we hold a short position, is a call option with log-moneyness $\log(K/S_0)=0.007$ and time-to-maturity $\tau_0=129$ calendar days. The option used for hedging (SPX211217C03950000) is a call with $\log(K/S_0)=0.01$ and higher time-to-expiry, $\tilde \tau_0=311$. The episode thus terminates when the main option has $\tau_n=44$ days left to expiration, which again is a major difference with respect to synthetic experiments. The time series of the three securities' price is displayed in the top-left panel (normalized such that $S_0=100$). The top-right and bottom-left panels show the different policies' positions in the S\&P 500 index and in the SPX option, respectively. In this example, PPO detaches from both the Delta-only and Delta-Vega practitioner hedging strategies. In particular, we observe that PPO over-hedges in the hedging option compared to Delta-Vega, and compensates by under-hedging the underlying asset, following the benchmarks' trajectory at a lower exposure. As we show below, this pattern is not specific to this episode.

\begin{table}[t]
    \centering
    \caption{%
    Sim-to-real hedging performance. Top: synthetic Heston test set, with the same calibrations, maturity buckets and episode structure as the training set. Bottom: full 2019--2022 S\&P 500 test set. Reported values are aggregated across four independent training runs.
    All quantities are in percent of $S_0$. 
    Bold marks the best (highest) mean-variance objective in each block.}
    \label{tab:sim2real}

    \begin{tabular}{@{}l
        S[table-format=-1.3] S[table-format=1.3]
        S[table-format=1.3] S[table-format=1.2]
        S[table-format=1.3] S[table-format=1.2]
        S[table-format=-2.2]
    @{}}
    \toprule
    \multirow{2}{*}{Strategy} & \multicolumn{2}{c}{Total P\&L} & \multicolumn{2}{c}{Asset market} & \multicolumn{2}{c}{Option market} & \multicolumn{1}{c}{\multirow{2}{*}{Objective}} \\
    \cmidrule(lr){2-3} \cmidrule(lr){4-5} \cmidrule(lr){6-7}
    & {Mean} & {Std} & {Cost} & {Turnover} & {Cost} & {Turnover} & \\
    \midrule
    \multicolumn{8}{@{}c}{\textit{Heston (n=10,000)}} \\[1pt]
    BS Delta    & -0.018 & 1.286 & 0.010 & 2.10 & 0.000 & 0.00 & -0.85 \\
    BS Delta-Vega & -0.032 & 0.582 & 0.006 & 0.92 & 0.020 & 1.73 & -0.20 \\
    RL         & -0.042 & 0.481 & 0.005 & 0.68 & 0.036 & 1.99 & $\bm{-0.16}$ \\
    \midrule

    \multicolumn{8}{@{}c}{\textit{S\&P 500, 2019--2022 (n=1193)}} \\[1pt]
    BS Delta   & -0.339 & 1.174 & 0.015 & 2.63 & 0.000 & 0.00 & -1.03 \\
    BS Delta-Vega & -0.047 & 0.560 & 0.010 & 1.40 & 0.050 & 1.95 & -0.20 \\
    RL         & 0.051  & 0.484 & 0.007 & 0.90 & 0.052 & 2.08 & $\bm{-0.07}$ \\
    \bottomrule
    \end{tabular}
\end{table}

Table \ref{tab:sim2real} reports the aggregate results of the sim-to-real transfer exercise, comparing the PPO agent to the two practitioner benchmarks on both the Heston test set (top block) and the S\&P 500 test set (bottom block). 
Overall, the results point to a successful transfer, as the agent outperforms the benchmarks in terms of mean-variance objective on the S\&P 500 data as it does on the simulated data.
We make three main observations.
First, the source of PPO's advantage over the practitioner benchmarks is different from the synthetic experiments of Section \ref{sec:res_costs}. There, the benchmark is the Heston model hedge, which is optimal (i.e., variance-minimizing) in the frictionless limit, so the agent could achieve a gain in the objective only by trading lower costs for higher P\&L dispersion. Here, the BS practitioner hedge is not variance-minimizing, neither under Heston nor on market data, so PPO achieves a better mean-variance objective mainly through a lower P\&L dispersion, with higher total costs on the Heston paths ($0.041\%$ versus Delta-Vega's $0.026\%$) and comparable costs on S\&P 500 data ($0.059\%$ versus $0.060\%$).
The P\&L standard deviation figures in Table \ref{tab:sim2real} also confirm that the advantage of hedging with two instruments is preserved on real data: both Delta-Vega and PPO reduce the standard deviation of the total P\&L by $52\%$ and $59\%$, respectively, compared to Delta-only, which is similar to the $55\%$ and $63\%$ figures observed on the held-out Heston paths.

Second, as in any sim-to-real transfer exercise, it is important to consider the differences between the synthetic training data and the empirical data used for testing. In training, the agent is exposed to paths generated under the risk-neutral measure $\mathbb{Q}$, which means that it cannot learn to profit from drift or risk premia. This is reflected in the Heston test, where the mean P\&L of $-0.042\%$ is approximately zero before costs ($0.041\%$ combining both instruments). In contrast, S\&P 500 price paths reflect real-world dynamics that include drifts and risk premia, which can lead to non-zero mean P\&L on market data, as we observe in the average total P\&L of PPO being positive ($0.051\%$). 

Third, and connected to the previous point, 
the size of the test set does not allow for strong statistical claims, since the mean P\&L estimates are not distinguishable from zero considering the 15 independent time windows as the effective sample size. We therefore read the mean-variance objective figures with caution, and focus on P\&L dispersion and on how the strategies allocate their hedge.

The hedging positions show a consistent pattern. Across both simulated and empirical test sets, PPO displays a systematic over-exposure to the hedging option and an under-exposure to the underlying asset compared to the practitioner Delta-Vega benchmark. Specifically, we find that PPO holds on average $0.77$ units of the hedging option on the S\&P 500 test set, against $0.70$ for Delta-Vega, and $0.12$ units of the underlying asset, against the benchmark's $0.18$. This is also observed in the simulated data ($0.76$ vs.\ $0.70$ for the hedging option, $0.11$ vs.\ $0.16$ for the underlying asset). 

This pattern is consistent with the Heston model on which the agent has been trained. As shown in Section \ref{sec:pract_hedge}, the model hedge differs from the practitioner one through the corrective factors $\chi_t^\nu, \chi_t^S$ in Equation \eqref{eq:heston_pract_hedge_comparison}. On our Heston test set, the model hedge holds on average $0.96$ units of the hedging option (after clipping to the agent's bounds), against $0.70$ for the practitioner hedge, thus indicating a correcting factor $\chi_t^\nu > 1$ on average. 
This is coherent with the Heston model's instantaneous variance following a CIR mean-reverting process (Equation \eqref{eq:heston_dv}), whereby a change in the current variance moves the implied volatility of the shorter-dated option more than that of the longer-dated hedging option, i.e., $\partial \sigma^{\rm imp}_t/\partial \nu_t \geq \partial \tilde \sigma^{\rm imp}_t / \partial \nu_t$. PPO thus learns a policy which is consistent with the data generating process of its training set (detaching from the pure model hedge to account for transaction costs, as shown in Section \ref{sec:res_costs}) and carries it to the market data.

\begin{figure}
    \centering
    \includegraphics[width=1\linewidth]{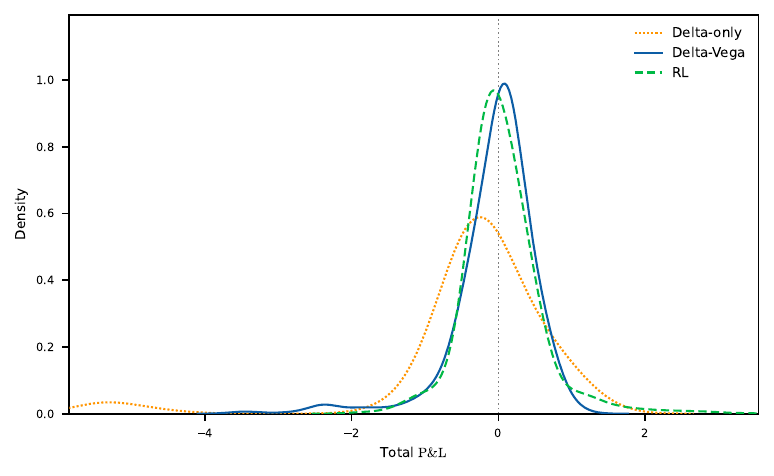}
    \caption{Total profit-and-Loss (P\&L) distribution for RL (PPO), Delta-only, and Delta-Vega hedging strategies on 2019-2022 SPX data under transaction costs $c=1$ BPS, $\tilde c=1\%$ (risk-aversion $\lambda=0.5$). 
    Curves are Kernel Density Estimates over a $1{,}193$ episodes test set.}
    \label{fig:hist_spx}
\end{figure}

Figure \ref{fig:hist_spx} shows the total P\&L distribution for the three hedging strategies over the S\&P 500 test set. PPO and Delta-Vega largely overlap, while Delta-only is more dispersed. The policies, however, differ in asymmetry and tail risk. The cumulative P\&L distributions of both benchmark strategies display negative skewness (Delta-Vega $-2.31$, Delta-only $-3.05$) and a 95\% Conditional Value-at-Risk ($\text{CVaR}_{95}$) of $1.84\%$ and $4.78\%$, respectively. PPO, on the other hand, yields a positively skewed P\&L (skewness $0.91$) with a lower tail risk ($\text{CVaR}_{95}=0.97\%$).
The episodes driving these statistics are mostly concentrated in 2020, when the S\&P 500 experienced a large drawdown and high volatility due to the pandemic shock. In particular, the 5\% worst episodes in terms of total P\&L for Delta-only all occur in 2020, while for Delta-Vega, 45 out of the 60 worst episodes fall in that year. PPO, on the other hand, has a less concentrated tail, with only about a third of its worst 5\% episodes occurring in 2020. 

We examine the regime dependence suggested by the tail analysis by reporting the year-by-year breakdown of the results in Table \ref{tab:sim2real_year}.
The pattern identified in the full set remains consistent across the years: we find that, on average, the agent holds more of the hedging option and less of the underlying asset than Delta-Vega in all years in our sample.
Furthermore, the size of the over-exposure compared to the practitioner strategy ranks with the single year's realized volatility. In 2020, the annualized standard deviation of daily log-returns of the underlying is $29\%$ (against a full sample figure of $19\%$), and PPO holds on average $20\%$ more of the hedging option than the practitioner policy.
In the same year, the net Vega exposure of PPO is largest (about $18\%$ of the main option's Vega), and so is its margin over the benchmarks.
This is also the year where the advantage of two-instrument hedging is most evident. Delta hedging posts a $2.18\%$ standard deviation of total P\&L, which is more than three times that of PPO ($0.72\%$) and more than double that of Delta-Vega ($1.03\%$). 

Outside 2020, the reduction in dispersion relative to Delta-only is smaller, about $28\%$ for Delta-Vega and $19\%$ for PPO. Accordingly, results are more balanced in the other years. In 2022, a year of high realized volatility ($24\%$), PPO ranks first, although with a higher P\&L standard deviation than Delta-Vega. In 2021, a calm year (realized volatility $13\%$), PPO underperforms Delta-Vega.
In 2019, also a stable year in terms of realized volatility, PPO's over-exposure to the hedging option is smallest (6\% more than Delta-Vega, on average),
and the agent achieves both lower total costs and a lower P\&L standard deviation.

\begin{table}[t]
    \centering
    \caption{%
    Sim-to-real hedging performance on the S\&P 500 test set, by year. Same conventions as Table~\ref{tab:sim2real}.}
    \label{tab:sim2real_year}

    \begin{tabular}{@{}l
        S[table-format=-1.3] S[table-format=1.3]
        S[table-format=1.3] S[table-format=1.2]
        S[table-format=1.3] S[table-format=1.2]
        S[table-format=-2.2]
    @{}}
    \toprule
    \multirow{2}{*}{Strategy} & \multicolumn{2}{c}{Total P\&L} & \multicolumn{2}{c}{Asset market} & \multicolumn{2}{c}{Option market} & \multicolumn{1}{c}{\multirow{2}{*}{Objective}} \\
    \cmidrule(lr){2-3} \cmidrule(lr){4-5} \cmidrule(lr){6-7}
    & {Mean} & {Std} & {Cost} & {Turnover} & {Cost} & {Turnover} & \\
    \midrule
    \multicolumn{8}{@{}c}{\textit{2019 (n=188)}} \\[1pt]
    BS Delta      & -0.178 & 0.377 & 0.019 & 3.01 & 0.000 & 0.00 & -0.25 \\
    BS Delta-Vega & -0.292 & 0.381 & 0.015 & 1.89 & 0.064 & 2.21 & -0.37 \\
    RL              & -0.151 & 0.306 & 0.008 & 1.02 & 0.041 & 2.01 & $\bm{-0.20}$ \\
    \midrule

    \multicolumn{8}{@{}c}{\textit{2020 (n=211)}} \\[1pt]
    BS Delta     & -1.565 & 2.182 & 0.018 & 3.04 & 0.000 & 0.00 & -3.94 \\
    BS Delta-Vega & -0.321 & 1.032 & 0.013 & 1.80 & 0.061 & 1.94 & -0.85 \\
    RL         & 0.072  & 0.718 & 0.008 & 1.00 & 0.061 & 2.21 & $\bm{-0.19}$ \\
    \midrule

    \multicolumn{8}{@{}c}{\textit{2021 (n=445)}} \\[1pt]
    BS Delta   & 0.149 & 0.581 & 0.012 & 2.32 & 0.000 & 0.00 & -0.02 \\
    BS Delta-Vega & 0.165  & 0.255 & 0.009 & 1.25 & 0.036 & 1.78 & $\bm{0.13}$ \\
    RL         & 0.087  & 0.378 & 0.007 & 0.93 & 0.046 & 1.99 & 0.02 \\
    \midrule

    \multicolumn{8}{@{}c}{\textit{2022 (n=349)}} \\[1pt]
    BS Delta    & -0.307 & 0.319 & 0.016 & 2.57 & 0.000 & 0.00 & -0.36 \\
    BS Delta-Vega & -0.018 & 0.365 & 0.009 & 1.08 & 0.054 & 2.02 & -0.08 \\
    RL         & 0.101  & 0.458 & 0.006 & 0.73 & 0.062 & 2.15 & $\bm{-0.00}$ \\
    \bottomrule
    \end{tabular}
\end{table}

We believe that our results provide an informative account of what a deep RL hedging policy trained in simulation and transferred to real market data can be expected to do in practice.
On the one hand, RL agents trained on risk-neutral price paths in the presence of market frictions can be successfully deployed to real market data, as PPO remains competitive with conventional benchmark strategies. 
On the other hand, the empirical experiments highlight how the hedging performance depends on the market simulator used for training, as we find that the agent acts on real data quite consistently with how it was trained in simulation.
As a result, the agent is likely to face the same shortcomings as the model it was trained on, which are reflected in residual risk exposures, hence on the realized P\&L. Nevertheless, compared to the model-based hedge, the agent still retains the advantage of naturally handling discrete time and nonlinearities such as transaction costs, which is arguably the main virtue of deep RL approach to hedging.
To conclude, we acknowledge that the size of our yearly samples does not support strong statistical claims. A higher frequency dataset or a panel of different underlying assets would allow a more rigorous assessment of the agent's robustness across market conditions.

\section{Conclusions and Future Research}\label{sec:conclusion}

We illustrate how a Proximal Policy Optimization (PPO) agent can be employed to hedge European options using two instruments simultaneously: the underlying asset, to manage price (Delta) risk, and a hedging option, to manage volatility (Vega) risk, under different transaction cost profiles.
We train the agent on a stochastic volatility environment simulated from the Heston model and we transfer the policy to real market data. 

We show that, in a frictionless Heston setting, PPO effectively replicates the theoretically optimal hedging policy. 
The agent recovers the semi-analytical benchmark from a parsimonious, market-observable state, without observing the latent instantaneous variance.

Once transaction costs are introduced, PPO learns to navigate the mean-variance trade-off between costs and replication errors. The agent outperforms the semi-analytical Delta-hedging and Delta-Vega benchmarks in mild and high friction scenarios, while achieving comparable results in the baseline low-cost calibration.
The mechanism is consistent across cost levels: PPO trades more conservatively than Delta-Vega, particularly in the option market, achieving lower transaction costs while keeping P\&L realized volatility from rising as sharply.

A sensitivity analysis over risk preferences confirms the robustness of this finding. PPO's transaction costs remain below those of Delta-Vega throughout the range we consider, and its P\&L volatility remains below that of Delta-only, occasionally falling below Delta-Vega's own level in higher friction, higher risk aversion settings.

We then conduct a sim-to-real transfer exercise, training the PPO agent under a distribution of calibrated Heston parameters and evaluating it against the practitioner Black-Scholes Delta and Delta-Vega hedging on daily S\&P 500 index option data from 2019 to 2022.
PPO and Delta-Vega roughly halve the P\&L dispersion of the single-instrument Delta hedge, 
and PPO attains a lower P\&L volatility than Delta-Vega at comparable transaction costs. PPO also attains the best mean-variance objective on the full empirical sample and in three of the four individual years, although the differences in mean P\&L are not statistically significant.
We find that PPO's advantage compared to the benchmarks is concentrated in episodes characterized by pronounced volatility.
This is consistent with the agent proposing a hedging strategy shaped by the Heston model on which it was trained. PPO holds more of the longer-dated option than the practitioner Delta-Vega, in line with the Heston model-implied hedge ratios, highlighting the importance of the training environment in shaping the learned policy.

Our work can be extended in a number of directions, which also address its main limitations.
The daily frequency and the limited temporal scope of the empirical dataset at hand limit the number of independent episodes available for testing, constraining the statistical strength of claims at a year level. A longer-horizon, higher-frequency dataset would allow a more rigorous assessment of out-of-sample performance across market regimes.

On the training side, the Heston model, while tractable and well-understood, is one  among many candidate simulators. Richer environments, such as rough or path-dependent volatility models, could improve the fidelity of the sim-to-real transfer. 
Similarly, hybrid approaches that combine synthetic and empirical data during training are already explored in the literature for single-instrument Delta hedging, and could be extended to a multi-instrument setting like the one studied in this work.

Our analysis is further confined to plain vanilla European options and a single mean-variance reward function. Extending the framework to exotic payoffs, for which hedging with multiple instruments is particularly important and tail-sensitive objectives may be preferable, is a natural next step.

Finally, the proportional cost model we adopt, while empirically grounded, abstracts from market impact and time-dependent spreads. Incorporating more realistic  microstructure, or directly considering observable bid-ask spreads, would bring the setting closer to the conditions faced by practitioners.

\newpage

\printbibliography

\newpage

\appendix

\section{Training Setup and Hyperparameters}
\label{app:hyper_train}

Table \ref{tab:hyperparams} reports the main hyperparameters for the synthetic (Section \ref{sec:res_costs}) and empirical (Section \ref{sec:res_simtoreal}) training runs. The two setups share the same PPO optimization hyperparameters but differ in network width, training length, and batch size. As mentioned in Section \ref{sec:emp_train}, this is to accommodate the larger and more heterogeneous dataset built for the sim-to-real exercise.

\begin{table}
\centering
\caption{Main PPO and environment hyperparameters for the synthetic and sim-to-real training runs.}
\label{tab:hyperparams}
\begin{tabular}{@{}l l l@{}}
\toprule
 & Synthetic & Sim-to-real \\
\midrule
\multicolumn{3}{@{}l}{\textit{Network architecture}} \\
Actor hidden layers      & $128\times128$ & $256\times256$ \\
Critic hidden layers     & $128\times128\times128\times128$ & $256\times256\times256\times256$ \\
\midrule
\multicolumn{3}{@{}l}{\textit{Optimization}} \\
Learning rate            & $3\times10^{-4}$ & $3\times10^{-4}$ \\
PPO clip ratio $\varepsilon$      & 0.1  & 0.1 \\
Critic loss weight $c_V$       & 0.5  & 0.5 \\
Discount $\gamma$        & 0.99 (0 if $c=\tilde c=0$) & 0.99 \\
$\lambda_{\text{GAE}}$ & 0.99 & 0.99 \\
Epochs per iteration  $N_\text{epochs}$   & 25   & 25 \\
Batch size   $N_\text{batch}$            & 128  & 256 \\
Training iterations $N_\text{iter}$     & 300  & 1{,}200 \\
\bottomrule
\end{tabular}
\end{table}

\begin{figure}
    \centering
    \includegraphics[width=.9\linewidth]{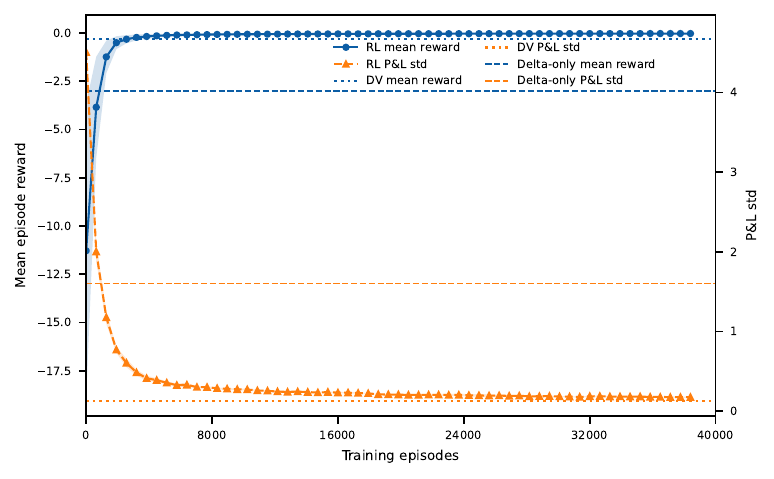}
    \caption{Mean episode reward (left axis) and P\&L standard deviation (right axis) on a held-out validation set, evaluated at each checkpoint during training in the frictionless Heston environment. Shaded bands show one standard deviation across four independent seeds. Dotted and dashed horizontal lines mark the metrics from the Heston model's Delta-Vega and Delta-only benchmark strategies.}
    \label{fig:train_frictionless}
\end{figure}

\begin{figure}
    \centering
    \includegraphics[width=.9\linewidth]{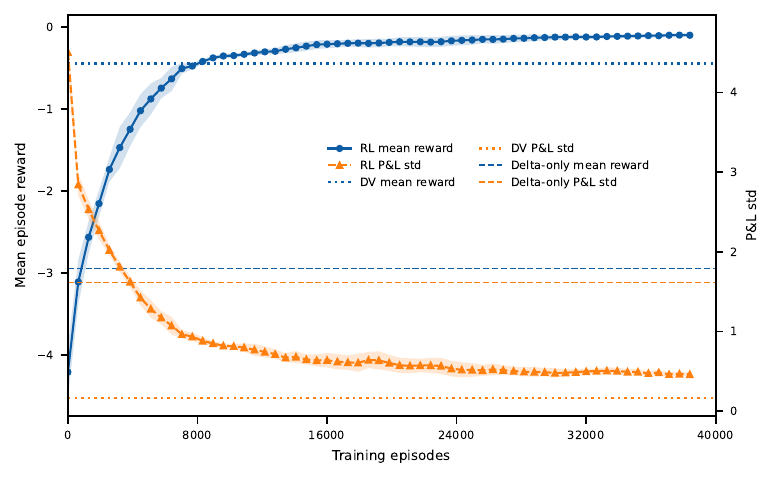}
    \caption{Mean episode reward (left axis) and P\&L standard deviation (right axis) on a held-out validation set, evaluated at each checkpoint during training in the Heston environment under transaction costs. Shaded bands show one standard deviation across four independent seeds. Dotted and dashed horizontal lines mark the metrics from the Heston model's Delta-Vega and Delta-only benchmark strategies.}
    \label{fig:train_costs}
\end{figure}

\begin{figure}
        \centering
        \includegraphics[width=.9\linewidth]{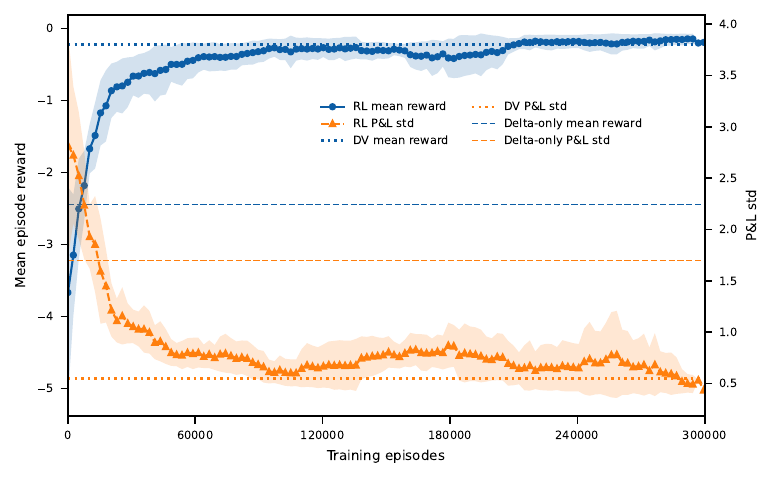}
        \caption{Mean episode reward (left axis) and P\&L standard deviation (right axis) on a held-out validation set, at different checkpoints during training on the multi-maturity, multi-calibration Heston dataset designed for sim-to-real. Shaded bands show one standard deviation across four independent seeds. Dotted and dashed horizontal lines mark the metrics from the BS practitioner Delta-Vega and Delta-only benchmark strategies.}
        \label{fig:train_emp}
    \end{figure}

Figure \ref{fig:train_frictionless} shows training curves for the frictionless Heston case averaged across four independent training runs. At different checkpoints during training, we evaluate the current policy by computing the mean episode reward and the cumulative P\&L standard deviation on a held-out validation set, whose size is equal to $N_{\text{batch}}$. Convergence is remarkably quick in this setting, with PPO reaching the Delta-Vega optimal policy's reward within the first training iterations. The total P\&L volatility over the validation set episodes reaches the benchmark's level by training end, accordingly.

Figure \ref{fig:train_costs} displays training metrics for the Heston environment with transaction costs (assuming the baseline calibration of Section \ref{sec:res_costs}). Differently from the frictionless case, PPO, although displaying a slower learning pattern at first, exceeds the benchmark reward level, which is not optimal anymore. We point out that there is no inconsistency between the mean reward shown here (higher for PPO than for the benchmark) and the objective values displayed in Table \ref{tab:inmodel_results_60d_spx_short_nonhp_agg_ra1}, which were essentially overlapping. As explained in Section \ref{sec:reward}, the reward function is an approximation of the objective value; furthermore, the test results reported in the table were computed on a $10{,}000$-episodes set, while validation metrics reported in the current figure are based on a $N_{\text{batch}}=128$-episodes set. 

Figure \ref{fig:train_emp} shows training curves for the sim-to-real runs.
Unlike the synthetic environments above, which are calibrated to a single option maturity and a single Heston parameterization, the current training set includes multiple maturities and calibrations (Section \ref{sec:emp_train}), yielding a harder and visibly noisier learning problem. We note that individual seeds occasionally display transient dips in validation reward during training. Because we apply early stopping with best-checkpoint restoration throughout, we report each seed's curve only up to its own best validation reward, i.e., the training trajectory of the policy actually retained and evaluated in Section \ref{sec:res_emptests}, rather than the full, noisier run. 
PPO reaches and eventually overtakes the validation reward secured by the BS practitioner Delta-Vega strategy. 
As already mentioned above, we stress that the curves being shown are reward over a $N_{\text{batch}}=256$-episodes validation set, which is meant to assess training health, while the final results in Section \ref{sec:res_emptests} display the actual mean-variance objective computed on $N_{\text{test}}=10,000$ synthetic episodes, as well as on $1{,}193$ historical episodes of S\&P 500 index and SPX options.

\end{document}